\documentclass[amsmath,amssymb,aps,twocolumn,showkeys,prl,floatfix]{revtex4}

\usepackage{mathrsfs}
\usepackage{amsmath}
\usepackage{amssymb}
\usepackage{color}
\usepackage{graphicx}	
\usepackage{bm}
\usepackage{ulem} 
\usepackage{titlesec}
\titleformat{\paragraph}[runin]
{\bfseries\scshape}{\theparagraph}{1em}{}
\usepackage{braket}
\usepackage{multirow}

\newcommand{\be}{\begin{equation}}
\newcommand{\ee}{\end{equation}}
\newcommand{\bef}{\begin{figure}}
\newcommand{\eef}{\end{figure}}
\newcommand{\bea}{\begin{eqnarray}}
\newcommand{\eea}{\end{eqnarray}}

\newcommand{\brho}{\mbox{\boldmath${\rho}$}}  
\DeclareMathOperator\erf{erf}
	     
\begin{document}
\title{Control of DNA denaturation bubble nucleation to advance nano-biosensing}
\author{Fran\c cois Sicard$^{1}$}
\thanks{Corresponding author: \texttt{francois.sicard@free.fr}.}
\author{Nicolas Destainville$^{2}$}
\author{Philippe Rousseau$^{3}$}
\author{Catherine Tardin$^{4}$}
\author{Manoel Manghi$^{2}$}
\affiliation{$^1$ Department of Chemical Engineering, University College London, United Kingdom}
\affiliation{$^2$ Laboratoire de Physique Th\'eorique, IRSAMC, Universit\'e de Toulouse, CNRS, UPS, France}
\affiliation{$^3$ Laboratoire de Microbiologie et G\'enetique Mol\'eculaires, Centre de Biologie Intégrative (CBI), Universit\'e de Toulouse, CNRS, UPS, France}
\affiliation{$^4$ Institut de Pharmacologie et Biologie Structurale, Universit\'e de Toulouse, CNRS, UPS, France}

\date{\today}
\begin{abstract}
In the demanding biosensing environment, improving selection efficiency strategies has become an issue of great significance.  
DNA minicircles containing between $200$ and $400$ base-pairs, also named microDNA, 
are representative of the supercoiled DNA loops found in nature. Their short size makes them extremely 
susceptible to writhe and twist, which is known to play a central role in DNA denaturation. 
We investigate minicircle lengths and superhelical densities that induce DNA denaturation bubbles 
of nanometer size and control well-defined long-life. Mesoscopic modeling and accelerated dynamics simulations 
allow us to study accurately the thermodynamic and dynamical properties associated with the nucleation 
and closure mechanisms of \textit{long-lived} denaturation bubbles.
Our results pave the way for new types of DNA biosensors with enhanced selectivity for specific DNA binding proteins. 
\end{abstract}


\maketitle

%
Biosensors, \textit{i.e.} analytical devices employing biological recognition properties 
for a selective bio-analysis~\cite{2016-FBB-Vigneshvar-Prakash}, have become very popular in recent years 
owning to their wide range of applications including clinical~\cite{2007-STJ-Anjum-Pundir}, 
environmental~\cite{2001-PAC-Maseini} and food analysis~\cite{1988-JFP-Eden-Schaertel}.
Since the invention of the first glucometer by Clark and Lyons~\cite{1962-ANYAS-Clark-Lyons}, 
biosensors have been developed for many different analytes, which range in size from 
individual ions~\cite{2013-EC-Han-Chen} to bacteria~\cite{2014-CMR-Ahmed-Millner}.
Due to their wide range of physical, chemical and biological activities, nucleic acid based biosensors have become 
increasingly important for rapid genetic screening and detection~\cite{2015-NAR-Brunet-Tardin,2017-JBBS-Kavita}. 
DNA interactions with proteins present specific challenges, such as the detection and measure of 
the levels of specific proteins in biological and environmental samples. As their detection, 
identification and quantification can be very complex, expensive and time 
consuming, the selection of highly efficient sensors is now required~\cite{2011-JAA-Ritzefeld-Sewald,
2012-CPC-Crawford-Kapanidis,2014-BB-Zhou-Ai}.

Both chemical and mechanical properties of the three dimensional structure of the DNA double helix 
have been examined to decipher the activity of specific target proteins~\cite{2012-CPC-Crawford-Kapanidis,2014-BB-Zhou-Ai}. 
Although the DNA macromolecule manifests more thermally driven opening of consecutive base-pairs (bps), 
also named  \textit{breathing} fluctuations, at physiological temperatures~\cite{2013-BP-Hippel-Marcus}, 
duplex opening can also be at play when non-linear elastic properties of DNA are involved. This commonly happens 
when the molecule is strongly bent~\cite{2009-BJ-Destainville-Palmeri} or negatively supercoiled~\cite{2012-SM-Adamcik-Dietler}. 
Various experimental~\cite{2003-PRL-Altan-Krichevsky} and analytical~\cite{2005-JPCM-Ambjornsson-Metzler,
2009-JPCM-Metzler-Fogedby,2015-JCP-Sicard-Manghi} models have been proposed in the literature to account for 
the thermodynamic and dynamical properties of denaturation bubbles. 
Yet this mechanical property of DNA has remained underused in the biosensing framework. Shi et al have recently 
taken advantage of  the existence of small breathing bubbles to induce isothermal polymerase chain reaction~\cite{2016-CC-Shi-Ma}. 
However, \textit{long-lived} denaturation bubbles extending over more than $4$ bps have not yet given rise 
to any biosensing application. This is largely due to poor knowledge of their properties. 
\begin{figure*}[t]
\includegraphics[width=0.95 \textwidth, angle=-0]{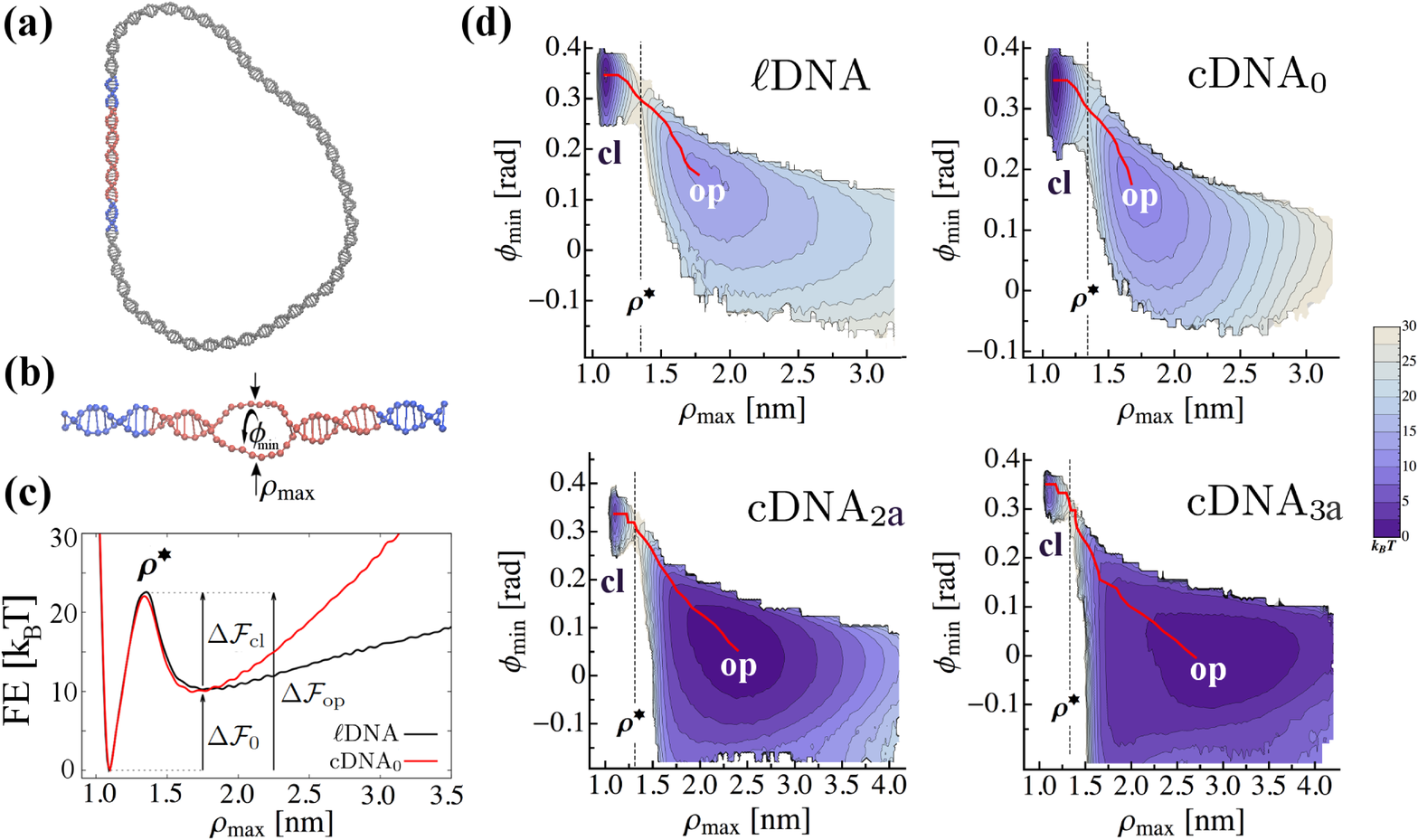}
\caption{
 Equilibrium snapshots of (a) circular DNA with pitch $p = 12.0$ bps ($\textrm{cDNA}_0$)  
 and (b) linear dsDNA ($\ell$DNA) when the \textit{long-lived} denaturation bubble is formed.
 The AT-rich region of size $30$ bps (red) is delimited at each extremity by two sequences 
 of 10 GC bps aligned arbitrarily along the Z-axis (blue). $\textrm{cDNA}_0$ is closed by a circular GC 
 region (grey). The maximal distance between paired bases, $\rho_{\textrm{max}}$, 
 and the minimal twist angle between successive bps, $\phi_{\textrm{min}}$, defined in the main text are shown.
(c) Free energy profiles associated with the opening/closure mechanism of $\ell$DNA and $\textrm{cDNA}_0$ projected 
along $\rho_{\textrm{max}}$. 
(d) Free energy surfaces projected along $\rho_{\textrm{max}}$ and $\phi_{\textrm{min}}$ in the linear 
and circular DNAs reported in Tab.~\ref{DNA-FES}. The free energy basins associated with the opened (op) and closed (cl) 
states of the DNA bubble and the typical minimal free energy paths obtained within the steepest descent framework (red) are shown.
}
\label{fig1}
\end{figure*}
\\

Here we elucidate the key parameters to obtain \textit{long-lived} bubbles at room temperature 
and we show how both their thermodynamic and dynamical properties could be worthwhile for various types of biosensors.
We consider DNA minicircles containing between $200$ and $400$ bps, 
also named microDNA, as they are representative of the supercoiled DNA loops found 
in nature~\cite{2012-Science-Shibata-Dutta} and have a suitable size for exploring 
the relationship between twist and writhe~\cite{2015-NC-Irobalieva-Zechiedrich}.
To overcome the inherent limitations of atomistic simulations encountered at length- and time-scales 
of interest~\cite{2016-PR-Manghi-Destainville}, mesoscopic modeling~\cite{2015-JCP-Sicard-Manghi} 
is combined with accelerated dynamics simulations~\cite{2013-PRL-Tiwary-Parrinello,2018-arXiv-Sicard} 
to study accurately the free energy landscape and the equilibrium rates associated with 
the nucleation/closure mechanisms of the \textit{long-lived} denaturation bubbles. 
We show how specific tuning of DNA structural parameters, such as the size and degree of supercoiling  can lead to 
a large variety of equilibrium closure/nucleation rates that can be seen as \textit{dynamical bandwidth} 
to advance the specificity of the biosensing probe and to reduce further the experimental setup complexity.\\

The double-stranded DNA (dsDNA) minicircle~\cite{2005-JCP-Mielke-Benham} is described at a mesoscopic scale, 
where the two single strands are modeled as freely rotating chains of $N$ beads 
of diameter $a=0.34$ nm with a AT-rich region of $30$ bps clamped by a closed circular GC region of 
$(N-30)$ bps~\cite{2015-JCP-Sicard-Manghi}. 
As shown in the Supporting Information (SI), the size of these AT-rich regions was chosen 
so that it is larger than the size of the representative \textit{long-lived} denaturation bubbles studied in this work.
We constrained a sequence of 10 GC bps on each extremity of the AT-rich region to be aligned arbitrarily 
along the Z-axis, as depicted in Fig.~\ref{fig1}~(a).
This allowed us to dissociate, in a first instance, the bending and twist contributions in the nucleation 
and closure mechanisms of the \textit{long-lived} denaturation bubble. 
The full Hamiltonian and the details of the numerical implementation and of the parameter values are given 
in previous works~\cite{2015-JCP-Sicard-Manghi} and in the SI.
The mesoscopic model yields numerical values for the dsDNA persistence length, $\ell_{\rm ds} \approx 160$~bps,
and the \textit{unconstrained} pitch, $p_0 = 12$~bps, comparable to the actual dsDNA values under 
physiological conditions~\cite{2015-CambridgeUniversityPress-Vologodskii}. 

In the following, we focused our analysis on one linear dsDNA ($\ell\textrm{DNA}$) of $N=50$ bps made of a AT-rich region 
of $30$ bps clamped by GC regions of $10$ bps on each extremity, and four different circular dsDNA (cDNA) 
with a similar AT-rich region but with different lengths and different superhelical densities, $\sigma$, 
defined as~\cite{2010-PRE-Sayar-etal}
\begin{equation}
\sigma = \frac{Lk - Lk^0}{Lk^0} = \frac{\Delta Lk}{Lk^0}  \,,
\label{superhelicaldensity}
\end{equation}
In Eq.~\ref{superhelicaldensity}, $Lk$ represent the linking numbers of the cDNA molecule~\cite{2010-PRE-Sayar-etal}, 
\textit{i.e.} the number of times one backbone strand \textit{links through} the circle formed by the other, 
and $Lk^0$ is defined as $Lk^0 = N/p_0$, for any DNA molecule, with $p_0 = 12.0$ (in bp units in the following) 
the equilibrium pitch measured in the linear state.
For a given molecule, the superhelical stress is accomodated by changes in helical twist, $\Delta Tw$, and writhe, $\Delta Wr$, 
following 
$\Delta Lk = \Delta Tw + \Delta Wr$~\cite{2005-JCP-Mielke-Benham}.

As shown in Table~\ref{DNA-FES}, we considered different values for $\sigma \in [-0.04;0]$.
For instance, natural circular DNA molecules, such as bacterial plasmids, vary widely in size, but, when isolated 
\textit{in vitro}, the majority have values for $ \sigma \leq -0.03$~\cite{2015-MS-Higgins-Vologodskii}. 
In the following, the superhelical densities, along with the sizes $N$ (in bp units) of the minicircles, were specifically chosen 
to tune the value of $\Delta Lk < 1$. Such specific design allowed us to control the interplay between twist and writhe 
during the formation of the \textit{long-lived} denaturation bubbles.\\
%

%
In Fig.~\ref{fig1}~(c) are shown the free energy profiles obtained within the metadynamics 
(metaD) framework~\cite{2013-PRL-Tiwary-Parrinello}, $\mathcal{F}$, 
associated with the nucleation and closure mechanisms  along the width $\rho_{\textrm{max}}$ of the bubble 
depicted in Fig.~\ref{fig1}~(b) for the linear and circular dsDNA with $\sigma = 0$ (cf. details in the SI).
In both systems, a \textit{closure} free energy barrier, $\Delta \mathcal{F}_{\textrm{cl}}$ $\approx 12.3~k_B T$ 
(with $T = 300$ K is room temperature) separates the metastable basins associated with the denaturation bubble 
($\rho_{\textrm{max}} \geq 1.35$ nm) from the closed state basin ($\rho_{\textrm{max}} \approx 1.1$ nm). 
These two basins are well separated by a standard free energy of formation $\Delta \mathcal{F}_0$ $\approx 10.3~k_B T$, 
defining the \textit{opening} free energy barrier, 
$\Delta \mathcal{F}_{\textrm{op}} \equiv \Delta \mathcal{F}_0 +  \Delta \mathcal{F}_{\textrm{cl}}$ $\approx 22.6~k_B T$, 
associated with the nucleation mechanism. 
These values can be compared with previous work~\cite{2015-JCP-Sicard-Manghi}, where the formation of denaturation bubble 
in linear dsDNA without restraint on the GC segments clamping the AT-rich region was studied. We measured a very 
similar value for $\Delta \mathcal{F}_{\textrm{op}}$, but a free energy difference 
of $\approx 2~k_B T$ in $\Delta \mathcal{F}_0$ ($8~k_B T$) and $\Delta \mathcal{F}_{\textrm{cl}}$ ($14~k_B T$). 
This difference in the free energies is about the thermal fluctuation scale and represents 
the loss of configurational entropy associated with the alignment of the GC regions during 
the closure of the AT-rich region.

We specifically designed $\textrm{cDNA}_0$ and $\ell\textrm{DNA}$ so that they differ from each other in terms 
of \textit{boundary} conditions with or without the closure of the GC regions located on each side 
of the AT-rich region. The closure condition yields the reduction of the configurational entropy 
contribution of the system in the metastable basin 
associated with the \textit{long-lived} denaturation bubble. This is qualitatively shown  
in Fig.~\ref{fig1}~(d) where the free energy surfaces (FES) are reconstructed within the metaD framework 
along the two collective variables (CVs), $\rho_{\textrm{max}}$ and $\phi_{\textrm{min}} = 
\textrm{min}_{i \in \textrm{bubble}}\phi_i$, depicted in Fig.~\ref{fig1}~(b). 
The entropic contribution to the FES can be quantitatively assessed considering the definition of the free energy difference 
in terms of the probability distribution of the CVs~\cite{2017-JCP-Gimondi-Salvalaglio,2018-arXiv-Sicard}, 
$\Delta F^{*}_{ij} = -k_B T ~\log \Big( \frac{P_i}{P_j}\Big)$, where $P_i$ and $P_j$ are the probabilities 
of states $i$ and $j$, respectively. The probability of each state is computed as 
\begin{equation}
 P_i = \iint_{(\rho_{\textrm{max}},\phi_{\textrm{min}}) \in \mathcal{B}_\textrm{i}} 
 f(\rho_{\textrm{max}},\phi_{\textrm{min}}) ~d\rho_{\textrm{max}} ~d\phi_{\textrm{min}} \,,
 \label{Proba-FE}
\end{equation}
where $f$ is the joint probability density distribution function associated with the system free energy.
The integration domains, $\mathcal{B}_\textrm{i}$, in Eq.~\ref{Proba-FE} are identified in the SI. We reported in Tab.~\ref{DNA-FES}.
the value of the free energy of formation, $\Delta F^{*}_0$, between the two basins observed in Fig.~\ref{fig1}~(d). 
As we could expect from visual inspection in Fig.~\ref{fig1}~(c)-(d), the free energy landscapes show significant 
differences between $\textrm{cDNA}_0$ and $\ell\textrm{DNA}$ associated with the difference in the \textit{global} 
entropic contribution in the free energy basins. 
\begin{table*}[t]
\begin{center}
\caption{Linear ($\ell$DNA) and circular (cDNA) DNA thermodynamic and dynamical properties obtained within 
the accelerated dynamics framework. $N/\ell_{\rm ds}$, $\sigma$, and $\Delta Lk$ correspond to the length, 
superhelical density and excess linking number of the molecules, respectively, and $\ell_{\rm ds}=160$ bps.
$\Delta F_0$, $\Delta F_{op}$, and $\Delta F_{cl}$ represent the algebric values of the free energy of formation, 
opening, and closure, respectively, measured along the minimal free energy paths depicted 
in Fig.~\ref{fig1}~(d). $\Delta F^*_0$ is the algebric values of the free energy 
of formation accounting for the entropic contribution in the system.
$\tau_{op}$ and $\tau_{cl}$ correspond to the characteristic times for the opening and closure 
of the \textit{long-lived} denaturation bubble.
}
\begin{tabular*}{0.95\textwidth}{@{\extracolsep{\fill}}cccccccccc}
  \hline\hline
  {} & $N/\ell_{\rm ds}$ & $\sigma$ & $\Delta Lk$ & $\Delta F_0$ ($k_B T$) & $\Delta F^*_0$ ($k_B T$) & $\Delta F_{op}$ ($k_B T$) & $\Delta F_{cl}$ ($k_B T$) & $\tau_{op}$ & $\tau_{cl}$ \\
  \hline \hline
  $\ell\textrm{DNA}$ & --- & --- & $0$ & $9.0 \pm 0.1$ & $6.7 \pm 0.1$ & $21.8 \pm 0.1$ & $12.9 \pm 0.1$ & $(67 \pm 8)$ ms & $(121 \pm 12)~\mu$s \\
  $\textrm{cDNA}_0$ & $1.9$ & $0$ & $0$ & $9.9 \pm 0.2$ & $8.1 \pm 0.2$ & $20.9 \pm 0.1$ & $11.1 \pm 0.2$ & $(51 \pm 3)$ ms & $(17 \pm 2)~\mu$s \\
   \hline
   $\textrm{cDNA}_{1a}$ & $1.5$ & $-0.008$ & $-0.17$ & $4.3 \pm 0.2$ & $1.2 \pm 0.2$ & $20.5 \pm 0.1$ & $15.9 \pm 0.2$ & $(10.4 \pm 0.6)$ ms& $(1.7 \pm 0.3)$ ms\\
   $\textrm{cDNA}_{1b}$ & $2.3$ & $-0.008$ & $-0.25$ & $6.5 \pm 0.2$ & $3.4 \pm 0.2$ & $21.0 \pm 0.1$ & $14.6 \pm 0.2$ & $(16.5 \pm 0.7)$ ms & $(0.33 \pm 0.02)$ ms \\
   \hline
  $\textrm{cDNA}_{2a}$ & $1.5$ & $-0.024$ & $-0.5$ & $-4.2 \pm 0.2$ & $-8.5 \pm 0.2$ & $19.7 \pm 0.2$ & $23.8 \pm 0.3$ & $(4.9 \pm 0.6)$ ms & $(90 \pm 30)$ min\\
  $\textrm{cDNA}_{2b}$ & $2.3$ & $-0.024$ & $-0.75$ & $-0.4 \pm 0.2$ & $-4.2 \pm 0.4$ & $21.6 \pm 0.1$ & $21.7 \pm 0.3$& $(5.9 \pm 0.5)$ ms& $(110 \pm 90)$ s\\
   \hline
  $\textrm{cDNA}_{3a}$ & $1.4$ & $-0.04$ & $-0.75$ & $-5.0 \pm 0.4$ & $-9.4 \pm 0.4$ & $21.8 \pm 0.1$ & $26.8 \pm 0.5$ & $(7.2 \pm 0.6)$ ms& $(6.8 \pm 3.2)$ h\\
  $\textrm{cDNA}_{3b}$ & $1.6$ & $-0.04$ & $-0.83$ & $-4.4 \pm 0.4$ & $-9.4 \pm 0.3$ & $20.9 \pm 0.3$ & $25.4 \pm 0.7$ & $(14.2 \pm 1.0)$ ms& $(22.5 \pm 9.0)$ h\\
  \hline
\end{tabular*}
\label{DNA-FES}
\end{center}
\end{table*}

In Fig.~\ref{fig1}~(d) is also shown the evolution of the FES, reconstructed within the metaD framework 
along $\rho_{\textrm{max}}$ and $\phi_{\textrm{min}}$, when the superhelical density of the system goes 
from $\sigma = 0$ to $-0.04$.
As we would expect from energetic consideration~\citep{2012-SM-Adamcik-Dietler}, we observe 
the progressive \textit{inversion} of the thermodynamic stability of the system 
for increasing undertwist, characteristic of the 
predominant stability of the \textit{long-lived} denaturation bubble. As shown in the SI, this transition comes with 
the drift of the location of the nucleation basin towards larger values of $\rho_{\textrm{max}}$, 
which is representative of the increase of the size of the denaturation bubble. 
As reported in Tab.~\ref{DNA-FES}, the impact of the superhelical density, $\sigma$, on the denaturation bubble 
stability is also shown with the increase of the \textit{closure} free energy, $\Delta F_{\textrm{cl}}$, measured along 
the minimal free energy paths (MFEPs) depicted in Fig.~\ref{fig1}~(d), which is maximal when $\sigma = -0.04$.\\

Interestingly, the results reported in Tab.~\ref{DNA-FES} show that the \textit{opening} free energy, 
$\Delta F_{\textrm{op}}$, measured along the MFEPs depicted in Fig.~\ref{fig1}~(d), 
does not significantly depends on the value of the superhelical density, $\sigma$. They suggest, however,
that the response of the cDNAs depends strongly on $N/\ell_{\rm ds}$ related to the flexibility of the dsDNAs.
This behavior is in line with the work of Sayar et al.~\cite{2010-PRE-Sayar-etal} where 
the fraction of the linking number absorbed as twist and writhe was studied when circular DNAs of different lengths 
approach the supercoiling transition. For dsDNA chains of the order of one persistence length, and  $\Delta Lk <1$,
the authors showed that the excess linking number was completely absorbed by the change in twist. 
For longer chains with $N/\ell_{\rm ds} >2$ (\textit{i.e.} longer than Kuhn's length in the dsDNA state), 
instead, they observed an increasing fraction of the linking number absorbed by the writhe. Indeed in this case 
the bending energy cost induced by the writhe is smaller.
In our cDNAs (cf. Tab.~\ref{DNA-FES}), this nontrivial dependence on chain length and excess linking number is reflected 
in the corresponding adjustment in the free energy of formation, $\Delta F^*_0$, and the 
\textit{closure} free energy, $\Delta F_{\textrm{cl}}$, measured along the MFEPs depicted in Fig.~\ref{fig1}~(d).

More sophisticated approach would necessarily take into account some relative misalignment of the sequences on both sides 
of the AT-rich region, at least during the initiation stage of the denaturation bubble nucleation/closure.
As we discussed quantitatively in the SI, the bending contribution can be assessed analytically by modeling 
the denaturation bubble as a single rotating joint, as the typical bubble length ($\sim 10$~bps) is 
on the order of the ssDNA persistence length, $\ell_p^{\rm ss} \simeq 4$~nm. As compared to the situation where 
the arms are forced to be aligned, we show that the free energy gain due to arm alignment 
is lower than $\sim 2.5 \, k_{\rm B}T$, in agreement with the result reported above for $\ell\textrm{DNA}$. \\

Finally, building on accelerated dynamics frameworks~\cite{2013-PRL-Tiwary-Parrinello} 
and the recent development of Sicard~\cite{2018-arXiv-Sicard} approaching the issue of complex system where configurational 
entropy is competing with energy, we assessed numerically 
the characteristic times associated with the opening and closure of the denaturation bubbles.  
The results reported in Tab.~\ref{DNA-FES} show a broad range of characteristic times associated 
either to the opening or the closure of the denaturation bubble of nanometer size (cf. details in the SI). 
For instance, the characteristic opening time and equilibrium constant obtained from our study in the case 
of the linear dsDNA ($\ell\textrm{DNA}$) are in good agreement with previous work~\cite{2015-JCP-Sicard-Manghi} 
and the experimental results of Englander et al.~\cite{1980-PNAS-Englander-Litwin} 
and more recently Altan-Bonnet et al.~\cite{2003-PRL-Altan-Krichevsky}.
As qualitatively shown in Fig.~\ref{fig1}~(d) and quantitatively assessed in the SI,
the results reported in Tab.~\ref{DNA-FES} show equilibrium times, which depend on the interplay between 
energetic and entropic characteristics of the undertwisted DNAs. For instance, we observed \textit{opening} 
times in the millisecond range, which are relatively unstressed by different degree of supercoiling. 
However, configurational entropy associated with the torsional constraint induced by similar $\sigma$ but different $\Delta Lk$ 
can significantly influence the \textit{closure} times over several orders of magnitude. \\

%
The extensive simulations discussed above allowed us to decipher the thermodynamic and dynamical characteristics 
of \textit{long-lived} nanometer-sized denaturation bubbles in \textit{undertwisted} microDNA containing between 200 and 400 bps.
Eventhough the numerical values derived above could be approximate because of our coarse-grained model, 
our results show that suitable tuning of the degree of supercoiling and size of specifically designed microDNA 
would allow the control of opening and closure characteristic times, ranging over well distinct timescales, 
from microseconds to several hours.
Interestingly, we showed that these dynamical characteristics can be related to specific 
tuning of both energetic and entropic properties of the DNA minicircles.

The broad range of closure/nucleation times could be seen as \textit{dynamical bandwidth} to advance 
the specificity of the biosensing probe. 
DNA supercoiling is determinant in the stability of these \textit{long-lived} DNA bubbles. 
The minicircles could therefore be used as a transducer of supercoiling induced by protein-binding resulting 
in bubble of various long lives. It could also permit to probe the DNA interaction of supercoiled-sensitive proteins 
with surface plasmon resonance technique~\cite{2017-PLoSONE-Pillet-Bouet} by enabling an easy immobilization of 
the minicircles through AT-rich ssDNA templates attached to the sensor surface.
Minicircles forming DNA bubbles with variable long lives could  also be used to detect and characterize 
the binding affinity of nucleoproteins for breathing DNA. A growing number of proteins implicated in fundamental 
biological processes such as transcription or repair are suspected to be extremely sensitive to such a DNA state. 
Alexandrov and coworkers reported a strong correlation between the binding affinity of the prokaryotic transcription 
factor Fis and enhanced breathing dynamics of the specific binding sequences~\cite{2013-PLOS-Nowak-Alexandrov}. 
The human Single-Stranded DNA binding protein 1 (hSSB1), involved in the repair of DNA damage, selectively counteract 
chemo- or radiotherapy cancer treatments, ensuring cancer cell survival~\cite{2015-NAR-Wu-Kang}. 
hSSB1 was shown to be recruited to dsDNA breaks within only 10~s after the breakage event as if hSSB1 had an enhanced 
sensitivity for breathing DNA~\cite{2018-SCDB-Croft-Richard}. The minicircles studied here could therefore permit 
to unravel the detailed mechanism of hSSB1 binding and its dynamics, and promote the design of new hSSB1 inhibitors, which would 
enhance the cell sensitivity to chemo-and radiotherapy and reduce the toxicity of anti-cancer-treatments. 
More generally, the biological mechanisms of single-stranded DNA binding proteins implicated in the maintenance 
of genome stability could largely benefit from the control of \textit{long-lived} nanometer-sized DNA denaturation bubbles 
forming in the minicircle explored here. 

%
We acknowledge L. Salom\'e and A.K. Dasanna for useful discussions. 
F.S. thanks J. Cuny and M. Salvalaglio for fruitful discussion concerning the metadynamics framework.

\bibliography{rsc} 

\begin{thebibliography}{35}
\expandafter\ifx\csname natexlab\endcsname\relax\def\natexlab#1{#1}\fi
\expandafter\ifx\csname bibnamefont\endcsname\relax
  \def\bibnamefont#1{#1}\fi
\expandafter\ifx\csname bibfnamefont\endcsname\relax
  \def\bibfnamefont#1{#1}\fi
\expandafter\ifx\csname citenamefont\endcsname\relax
  \def\citenamefont#1{#1}\fi
\expandafter\ifx\csname url\endcsname\relax
  \def\url#1{\texttt{#1}}\fi
\expandafter\ifx\csname urlprefix\endcsname\relax\def\urlprefix{URL }\fi
\providecommand{\bibinfo}[2]{#2}
\providecommand{\eprint}[2][]{\url{#2}}

\bibitem[{\citenamefont{Vigneshvar et~al.}(2016)\citenamefont{Vigneshvar,
  Sudhakumari, Senthilkumaran, and Prakash}}]{2016-FBB-Vigneshvar-Prakash}
\bibinfo{author}{\bibfnamefont{S.}~\bibnamefont{Vigneshvar}},
  \bibinfo{author}{\bibfnamefont{C.}~\bibnamefont{Sudhakumari}},
  \bibinfo{author}{\bibfnamefont{B.}~\bibnamefont{Senthilkumaran}},
  \bibnamefont{and} \bibinfo{author}{\bibfnamefont{H.}~\bibnamefont{Prakash}},
  \bibinfo{journal}{Front. Bioeng. Biotechnol.} \textbf{\bibinfo{volume}{4}},
  \bibinfo{pages}{11} (\bibinfo{year}{2016}).

\bibitem[{\citenamefont{Anjum and Pundir}(2007)}]{2007-STJ-Anjum-Pundir}
\bibinfo{author}{\bibfnamefont{V.}~\bibnamefont{Anjum}} \bibnamefont{and}
  \bibinfo{author}{\bibfnamefont{C.}~\bibnamefont{Pundir}},
  \bibinfo{journal}{Sensors and Tranducers Journal}
  \textbf{\bibinfo{volume}{76}}, \bibinfo{pages}{937} (\bibinfo{year}{2007}).

\bibitem[{\citenamefont{Maseini}(2001)}]{2001-PAC-Maseini}
\bibinfo{author}{\bibfnamefont{M.}~\bibnamefont{Maseini}},
  \bibinfo{journal}{Pure Appl. Chem.} \textbf{\bibinfo{volume}{73}},
  \bibinfo{pages}{23} (\bibinfo{year}{2001}).

\bibitem[{\citenamefont{Eden-Firstenberg and
  Schaertel}(1988)}]{1988-JFP-Eden-Schaertel}
\bibinfo{author}{\bibfnamefont{R.}~\bibnamefont{Eden-Firstenberg}}
  \bibnamefont{and}
  \bibinfo{author}{\bibfnamefont{B.}~\bibnamefont{Schaertel}},
  \bibinfo{journal}{J. of Food Protection} \textbf{\bibinfo{volume}{51}},
  \bibinfo{pages}{811} (\bibinfo{year}{1988}).

\bibitem[{\citenamefont{Jr. and Lyons}(1962)}]{1962-ANYAS-Clark-Lyons}
\bibinfo{author}{\bibfnamefont{L.~C.} \bibnamefont{Jr.}} \bibnamefont{and}
  \bibinfo{author}{\bibfnamefont{C.}~\bibnamefont{Lyons}},
  \bibinfo{journal}{Ann. N. Y. Acad. Sci.} \textbf{\bibinfo{volume}{102}},
  \bibinfo{pages}{29} (\bibinfo{year}{1962}).

\bibitem[{\citenamefont{Han et~al.}(2013)\citenamefont{Han, Ma, Zhao, Xu, and
  Chen}}]{2013-EC-Han-Chen}
\bibinfo{author}{\bibfnamefont{D.}~\bibnamefont{Han}},
  \bibinfo{author}{\bibfnamefont{Z.}~\bibnamefont{Ma}},
  \bibinfo{author}{\bibfnamefont{W.}~\bibnamefont{Zhao}},
  \bibinfo{author}{\bibfnamefont{J.}~\bibnamefont{Xu}}, \bibnamefont{and}
  \bibinfo{author}{\bibfnamefont{H.}~\bibnamefont{Chen}},
  \bibinfo{journal}{Electrochem. Commun.} \textbf{\bibinfo{volume}{35}},
  \bibinfo{pages}{38} (\bibinfo{year}{2013}).

\bibitem[{\citenamefont{Ahmed et~al.}(2014)\citenamefont{Ahmed, Rushworth,
  Hirst, and Millner}}]{2014-CMR-Ahmed-Millner}
\bibinfo{author}{\bibfnamefont{A.}~\bibnamefont{Ahmed}},
  \bibinfo{author}{\bibfnamefont{J.}~\bibnamefont{Rushworth}},
  \bibinfo{author}{\bibfnamefont{N.}~\bibnamefont{Hirst}}, \bibnamefont{and}
  \bibinfo{author}{\bibfnamefont{P.}~\bibnamefont{Millner}},
  \bibinfo{journal}{Clinical Microbiol. Rev.} \textbf{\bibinfo{volume}{27}},
  \bibinfo{pages}{631} (\bibinfo{year}{2014}).

\bibitem[{\citenamefont{Brunet et~al.}(2015)\citenamefont{Brunet, Chevalier,
  Destainville, Manghi, Rousseau, Salhi, Salom\'e, and
  Tardin}}]{2015-NAR-Brunet-Tardin}
\bibinfo{author}{\bibfnamefont{A.}~\bibnamefont{Brunet}},
  \bibinfo{author}{\bibfnamefont{S.}~\bibnamefont{Chevalier}},
  \bibinfo{author}{\bibfnamefont{N.}~\bibnamefont{Destainville}},
  \bibinfo{author}{\bibfnamefont{M.}~\bibnamefont{Manghi}},
  \bibinfo{author}{\bibfnamefont{P.}~\bibnamefont{Rousseau}},
  \bibinfo{author}{\bibfnamefont{M.}~\bibnamefont{Salhi}},
  \bibinfo{author}{\bibfnamefont{L.}~\bibnamefont{Salom\'e}}, \bibnamefont{and}
  \bibinfo{author}{\bibfnamefont{C.}~\bibnamefont{Tardin}},
  \bibinfo{journal}{Nucleic Acids Research} \textbf{\bibinfo{volume}{43}},
  \bibinfo{pages}{e72} (\bibinfo{year}{2015}).

\bibitem[{\citenamefont{Kavita}(2017)}]{2017-JBBS-Kavita}
\bibinfo{author}{\bibfnamefont{V.}~\bibnamefont{Kavita}}, \bibinfo{journal}{J.
  Bioengineer. Biomedical Sci.} \textbf{\bibinfo{volume}{7}},
  \bibinfo{pages}{222} (\bibinfo{year}{2017}).

\bibitem[{\citenamefont{Ritzefeld and
  Sewald}(2011)}]{2011-JAA-Ritzefeld-Sewald}
\bibinfo{author}{\bibfnamefont{M.}~\bibnamefont{Ritzefeld}} \bibnamefont{and}
  \bibinfo{author}{\bibfnamefont{N.}~\bibnamefont{Sewald}},
  \bibinfo{journal}{J. Amino Acids} \textbf{\bibinfo{volume}{2012}},
  \bibinfo{pages}{816032} (\bibinfo{year}{2011}).

\bibitem[{\citenamefont{Crawford et~al.}(2012)\citenamefont{Crawford, Kelly,
  and Kapanidis}}]{2012-CPC-Crawford-Kapanidis}
\bibinfo{author}{\bibfnamefont{R.}~\bibnamefont{Crawford}},
  \bibinfo{author}{\bibfnamefont{D.}~\bibnamefont{Kelly}}, \bibnamefont{and}
  \bibinfo{author}{\bibfnamefont{A.}~\bibnamefont{Kapanidis}},
  \bibinfo{journal}{Chem. Phys. Chem.} \textbf{\bibinfo{volume}{13}},
  \bibinfo{pages}{918} (\bibinfo{year}{2012}).

\bibitem[{\citenamefont{Zhou et~al.}(2014)\citenamefont{Zhou, Xu, Wang, Sun,
  Yin, and Ai}}]{2014-BB-Zhou-Ai}
\bibinfo{author}{\bibfnamefont{Y.}~\bibnamefont{Zhou}},
  \bibinfo{author}{\bibfnamefont{Z.}~\bibnamefont{Xu}},
  \bibinfo{author}{\bibfnamefont{M.}~\bibnamefont{Wang}},
  \bibinfo{author}{\bibfnamefont{B.}~\bibnamefont{Sun}},
  \bibinfo{author}{\bibfnamefont{H.}~\bibnamefont{Yin}}, \bibnamefont{and}
  \bibinfo{author}{\bibfnamefont{S.}~\bibnamefont{Ai}},
  \bibinfo{journal}{Biosens. Bioelectron.} \textbf{\bibinfo{volume}{53}},
  \bibinfo{pages}{263} (\bibinfo{year}{2014}).

\bibitem[{\citenamefont{von Hippel et~al.}(2013)\citenamefont{von Hippel,
  Johnson, and Marcus}}]{2013-BP-Hippel-Marcus}
\bibinfo{author}{\bibfnamefont{P.}~\bibnamefont{von Hippel}},
  \bibinfo{author}{\bibfnamefont{N.}~\bibnamefont{Johnson}}, \bibnamefont{and}
  \bibinfo{author}{\bibfnamefont{A.}~\bibnamefont{Marcus}},
  \bibinfo{journal}{Biopolymers} \textbf{\bibinfo{volume}{99}},
  \bibinfo{pages}{923} (\bibinfo{year}{2013}).

\bibitem[{\citenamefont{Destainville et~al.}(2009)\citenamefont{Destainville,
  Manghi, and Palmeri}}]{2009-BJ-Destainville-Palmeri}
\bibinfo{author}{\bibfnamefont{N.}~\bibnamefont{Destainville}},
  \bibinfo{author}{\bibfnamefont{M.}~\bibnamefont{Manghi}}, \bibnamefont{and}
  \bibinfo{author}{\bibfnamefont{J.}~\bibnamefont{Palmeri}},
  \bibinfo{journal}{Biophys. J.} \textbf{\bibinfo{volume}{96}},
  \bibinfo{pages}{4464} (\bibinfo{year}{2009}).

\bibitem[{\citenamefont{Adamcik et~al.}(2012)\citenamefont{Adamcik, Jeon,
  Karczewski, Metzler, and Dietler}}]{2012-SM-Adamcik-Dietler}
\bibinfo{author}{\bibfnamefont{J.}~\bibnamefont{Adamcik}},
  \bibinfo{author}{\bibfnamefont{J.-H.} \bibnamefont{Jeon}},
  \bibinfo{author}{\bibfnamefont{K.}~\bibnamefont{Karczewski}},
  \bibinfo{author}{\bibfnamefont{R.}~\bibnamefont{Metzler}}, \bibnamefont{and}
  \bibinfo{author}{\bibfnamefont{G.}~\bibnamefont{Dietler}},
  \bibinfo{journal}{Soft Matter} \textbf{\bibinfo{volume}{8}},
  \bibinfo{pages}{8651} (\bibinfo{year}{2012}).

\bibitem[{\citenamefont{Altan-Bonnet et~al.}(2003)\citenamefont{Altan-Bonnet,
  Libchaber, and Krichevsky}}]{2003-PRL-Altan-Krichevsky}
\bibinfo{author}{\bibfnamefont{G.}~\bibnamefont{Altan-Bonnet}},
  \bibinfo{author}{\bibfnamefont{A.}~\bibnamefont{Libchaber}},
  \bibnamefont{and}
  \bibinfo{author}{\bibfnamefont{O.}~\bibnamefont{Krichevsky}},
  \bibinfo{journal}{Phys. Rev. Lett.} \textbf{\bibinfo{volume}{90}},
  \bibinfo{pages}{138101} (\bibinfo{year}{2003}).

\bibitem[{\citenamefont{Ambjornsson and
  Metzler}(2005)}]{2005-JPCM-Ambjornsson-Metzler}
\bibinfo{author}{\bibfnamefont{T.}~\bibnamefont{Ambjornsson}} \bibnamefont{and}
  \bibinfo{author}{\bibfnamefont{R.}~\bibnamefont{Metzler}},
  \bibinfo{journal}{J. Phys.: Condens. Matter} \textbf{\bibinfo{volume}{17}},
  \bibinfo{pages}{S1841} (\bibinfo{year}{2005}).

\bibitem[{\citenamefont{Metzler et~al.}(2009)\citenamefont{Metzler,
  Ambjornsson, Hanke, and Fogedby}}]{2009-JPCM-Metzler-Fogedby}
\bibinfo{author}{\bibfnamefont{R.}~\bibnamefont{Metzler}},
  \bibinfo{author}{\bibfnamefont{T.}~\bibnamefont{Ambjornsson}},
  \bibinfo{author}{\bibfnamefont{A.}~\bibnamefont{Hanke}}, \bibnamefont{and}
  \bibinfo{author}{\bibfnamefont{H.}~\bibnamefont{Fogedby}},
  \bibinfo{journal}{J. Phys.: Condens. Matter} \textbf{\bibinfo{volume}{21}},
  \bibinfo{pages}{034111} (\bibinfo{year}{2009}).

\bibitem[{\citenamefont{Sicard et~al.}(2015)\citenamefont{Sicard, Destainville,
  and Manghi}}]{2015-JCP-Sicard-Manghi}
\bibinfo{author}{\bibfnamefont{F.}~\bibnamefont{Sicard}},
  \bibinfo{author}{\bibfnamefont{N.}~\bibnamefont{Destainville}},
  \bibnamefont{and} \bibinfo{author}{\bibfnamefont{M.}~\bibnamefont{Manghi}},
  \bibinfo{journal}{J. Chem. Phys.} \textbf{\bibinfo{volume}{142}},
  \bibinfo{pages}{034903} (\bibinfo{year}{2015}).

\bibitem[{\citenamefont{Shi et~al.}(2016)\citenamefont{Shi, Shang, Zhou, Zhang,
  Wang, and Ma}}]{2016-CC-Shi-Ma}
\bibinfo{author}{\bibfnamefont{C.}~\bibnamefont{Shi}},
  \bibinfo{author}{\bibfnamefont{F.}~\bibnamefont{Shang}},
  \bibinfo{author}{\bibfnamefont{M.}~\bibnamefont{Zhou}},
  \bibinfo{author}{\bibfnamefont{P.}~\bibnamefont{Zhang}},
  \bibinfo{author}{\bibfnamefont{Y.}~\bibnamefont{Wang}}, \bibnamefont{and}
  \bibinfo{author}{\bibfnamefont{C.}~\bibnamefont{Ma}}, \bibinfo{journal}{Chem.
  Commun.} \textbf{\bibinfo{volume}{52}}, \bibinfo{pages}{11551}
  (\bibinfo{year}{2016}).

\bibitem[{\citenamefont{Shibata et~al.}(2012)\citenamefont{Shibata, Kumar,
  Layer, Willcox, Gagan, Griffith, and Dutta}}]{2012-Science-Shibata-Dutta}
\bibinfo{author}{\bibfnamefont{Y.}~\bibnamefont{Shibata}},
  \bibinfo{author}{\bibfnamefont{P.}~\bibnamefont{Kumar}},
  \bibinfo{author}{\bibfnamefont{R.}~\bibnamefont{Layer}},
  \bibinfo{author}{\bibfnamefont{S.}~\bibnamefont{Willcox}},
  \bibinfo{author}{\bibfnamefont{J.}~\bibnamefont{Gagan}},
  \bibinfo{author}{\bibfnamefont{J.}~\bibnamefont{Griffith}}, \bibnamefont{and}
  \bibinfo{author}{\bibfnamefont{A.}~\bibnamefont{Dutta}},
  \bibinfo{journal}{Science} \textbf{\bibinfo{volume}{336}},
  \bibinfo{pages}{82} (\bibinfo{year}{2012}).

\bibitem[{\citenamefont{Irobalieva et~al.}(2015)\citenamefont{Irobalieva, Fogg,
  Jr., Sutthibutpong, Chen, Barker, Ludtke, Harris, Schmid, Chiu
  et~al.}}]{2015-NC-Irobalieva-Zechiedrich}
\bibinfo{author}{\bibfnamefont{R.}~\bibnamefont{Irobalieva}},
  \bibinfo{author}{\bibfnamefont{J.}~\bibnamefont{Fogg}},
  \bibinfo{author}{\bibfnamefont{D.~C.} \bibnamefont{Jr.}},
  \bibinfo{author}{\bibfnamefont{T.}~\bibnamefont{Sutthibutpong}},
  \bibinfo{author}{\bibfnamefont{M.}~\bibnamefont{Chen}},
  \bibinfo{author}{\bibfnamefont{A.}~\bibnamefont{Barker}},
  \bibinfo{author}{\bibfnamefont{S.}~\bibnamefont{Ludtke}},
  \bibinfo{author}{\bibfnamefont{S.}~\bibnamefont{Harris}},
  \bibinfo{author}{\bibfnamefont{M.}~\bibnamefont{Schmid}},
  \bibinfo{author}{\bibfnamefont{W.}~\bibnamefont{Chiu}}, \bibnamefont{et~al.},
  \bibinfo{journal}{Nature Comm.} \textbf{\bibinfo{volume}{6}},
  \bibinfo{pages}{8440} (\bibinfo{year}{2015}).

\bibitem[{\citenamefont{Manghi and
  Destainville}(2016)}]{2016-PR-Manghi-Destainville}
\bibinfo{author}{\bibfnamefont{M.}~\bibnamefont{Manghi}} \bibnamefont{and}
  \bibinfo{author}{\bibfnamefont{N.}~\bibnamefont{Destainville}},
  \bibinfo{journal}{Phys. Rep.} \textbf{\bibinfo{volume}{631}},
  \bibinfo{pages}{1} (\bibinfo{year}{2016}).

\bibitem[{\citenamefont{Tiwary and
  Parrinello}(2013)}]{2013-PRL-Tiwary-Parrinello}
\bibinfo{author}{\bibfnamefont{P.}~\bibnamefont{Tiwary}} \bibnamefont{and}
  \bibinfo{author}{\bibfnamefont{M.}~\bibnamefont{Parrinello}},
  \bibinfo{journal}{Phys. Rev. Lett.} \textbf{\bibinfo{volume}{111}},
  \bibinfo{pages}{230602} (\bibinfo{year}{2013}).

\bibitem[{\citenamefont{Sicard}(2018)}]{2018-arXiv-Sicard}
\bibinfo{author}{\bibfnamefont{F.}~\bibnamefont{Sicard}},
  \bibinfo{journal}{arXiv:1803.03490 [cond-mat.stat-mech]}
  (\bibinfo{year}{2018}).

\bibitem[{\citenamefont{Mielke et~al.}(2005)\citenamefont{Mielke,
  Gr\o{}nbech-Jensen, Krishnan, Fink, and Benham}}]{2005-JCP-Mielke-Benham}
\bibinfo{author}{\bibfnamefont{S.}~\bibnamefont{Mielke}},
  \bibinfo{author}{\bibfnamefont{N.}~\bibnamefont{Gr\o{}nbech-Jensen}},
  \bibinfo{author}{\bibfnamefont{V.}~\bibnamefont{Krishnan}},
  \bibinfo{author}{\bibfnamefont{W.}~\bibnamefont{Fink}}, \bibnamefont{and}
  \bibinfo{author}{\bibfnamefont{C.}~\bibnamefont{Benham}},
  \bibinfo{journal}{J. Chem. Phys.} \textbf{\bibinfo{volume}{123}},
  \bibinfo{pages}{124911} (\bibinfo{year}{2005}).

\bibitem[{\citenamefont{Vologodskii}(2015)}]{2015-CambridgeUniversityPress-Vologodskii}
\bibinfo{author}{\bibfnamefont{A.}~\bibnamefont{Vologodskii}},
  \emph{\bibinfo{title}{Biophysics of DNA}} (\bibinfo{publisher}{Cambridge
  University Press}, \bibinfo{year}{2015}).

\bibitem[{\citenamefont{Sayar et~al.}(2010)\citenamefont{Sayar, Avsaroglu, and
  Kabakcioglu}}]{2010-PRE-Sayar-etal}
\bibinfo{author}{\bibfnamefont{M.}~\bibnamefont{Sayar}},
  \bibinfo{author}{\bibfnamefont{B.}~\bibnamefont{Avsaroglu}},
  \bibnamefont{and}
  \bibinfo{author}{\bibfnamefont{A.}~\bibnamefont{Kabakcioglu}},
  \bibinfo{journal}{Phys. Rev. E} \textbf{\bibinfo{volume}{81}},
  \bibinfo{pages}{041916} (\bibinfo{year}{2010}).

\bibitem[{\citenamefont{Higgins and
  Vologodskii}(2015)}]{2015-MS-Higgins-Vologodskii}
\bibinfo{author}{\bibfnamefont{N.}~\bibnamefont{Higgins}} \bibnamefont{and}
  \bibinfo{author}{\bibfnamefont{A.}~\bibnamefont{Vologodskii}},
  \bibinfo{journal}{Microbiol. Spectr.} \textbf{\bibinfo{volume}{3}},
  \bibinfo{pages}{1} (\bibinfo{year}{2015}).

\bibitem[{\citenamefont{Gimondi and
  Salvalaglio}(2017)}]{2017-JCP-Gimondi-Salvalaglio}
\bibinfo{author}{\bibfnamefont{I.}~\bibnamefont{Gimondi}} \bibnamefont{and}
  \bibinfo{author}{\bibfnamefont{M.}~\bibnamefont{Salvalaglio}},
  \bibinfo{journal}{J. Chem. Phys.} \textbf{\bibinfo{volume}{147}},
  \bibinfo{pages}{114502} (\bibinfo{year}{2017}).

\bibitem[{\citenamefont{Englander et~al.}(1980)\citenamefont{Englander,
  Kallenbach, Heeger, Krumhansl, and Litwin}}]{1980-PNAS-Englander-Litwin}
\bibinfo{author}{\bibfnamefont{S.}~\bibnamefont{Englander}},
  \bibinfo{author}{\bibfnamefont{N.}~\bibnamefont{Kallenbach}},
  \bibinfo{author}{\bibfnamefont{A.}~\bibnamefont{Heeger}},
  \bibinfo{author}{\bibfnamefont{J.}~\bibnamefont{Krumhansl}},
  \bibnamefont{and} \bibinfo{author}{\bibfnamefont{S.}~\bibnamefont{Litwin}},
  \bibinfo{journal}{Proc. Nat. Acad. Soc. U.S.A.}
  \textbf{\bibinfo{volume}{77}}, \bibinfo{pages}{7222} (\bibinfo{year}{1980}).

\bibitem[{\citenamefont{Pillet et~al.}(2017)\citenamefont{Pillet, Passot,
  Pasta, Leberre, and Bouet}}]{2017-PLoSONE-Pillet-Bouet}
\bibinfo{author}{\bibfnamefont{F.}~\bibnamefont{Pillet}},
  \bibinfo{author}{\bibfnamefont{F.~M.} \bibnamefont{Passot}},
  \bibinfo{author}{\bibfnamefont{F.}~\bibnamefont{Pasta}},
  \bibinfo{author}{\bibfnamefont{V.~A.} \bibnamefont{Leberre}},
  \bibnamefont{and} \bibinfo{author}{\bibfnamefont{J.-Y.} \bibnamefont{Bouet}},
  \bibinfo{journal}{PLoS ONE} \textbf{\bibinfo{volume}{12}},
  \bibinfo{pages}{e0177056} (\bibinfo{year}{2017}).

\bibitem[{\citenamefont{Nowak-Lovato et~al.}(2013)\citenamefont{Nowak-Lovato,
  Alexandrov, Banisadr, Bauer, Bishop, Usheva, Mu, Hong-Geller, Rasmussen, and
  Hlavacek}}]{2013-PLOS-Nowak-Alexandrov}
\bibinfo{author}{\bibfnamefont{K.}~\bibnamefont{Nowak-Lovato}},
  \bibinfo{author}{\bibfnamefont{L.}~\bibnamefont{Alexandrov}},
  \bibinfo{author}{\bibfnamefont{A.}~\bibnamefont{Banisadr}},
  \bibinfo{author}{\bibfnamefont{A.}~\bibnamefont{Bauer}},
  \bibinfo{author}{\bibfnamefont{A.}~\bibnamefont{Bishop}},
  \bibinfo{author}{\bibfnamefont{A.}~\bibnamefont{Usheva}},
  \bibinfo{author}{\bibfnamefont{F.}~\bibnamefont{Mu}},
  \bibinfo{author}{\bibfnamefont{E.}~\bibnamefont{Hong-Geller}},
  \bibinfo{author}{\bibfnamefont{K.}~\bibnamefont{Rasmussen}},
  \bibnamefont{and} \bibinfo{author}{\bibfnamefont{W.}~\bibnamefont{Hlavacek}},
  \bibinfo{journal}{PLoS Comput. Biol.} \textbf{\bibinfo{volume}{9}},
  \bibinfo{pages}{e1002881} (\bibinfo{year}{2013}).

\bibitem[{\citenamefont{Wu et~al.}(2015)\citenamefont{Wu, Chen, Lu, Zhang,
  Zhang, Duan, Wang, Huang, and Kang}}]{2015-NAR-Wu-Kang}
\bibinfo{author}{\bibfnamefont{Y.}~\bibnamefont{Wu}},
  \bibinfo{author}{\bibfnamefont{H.}~\bibnamefont{Chen}},
  \bibinfo{author}{\bibfnamefont{J.}~\bibnamefont{Lu}},
  \bibinfo{author}{\bibfnamefont{M.}~\bibnamefont{Zhang}},
  \bibinfo{author}{\bibfnamefont{R.}~\bibnamefont{Zhang}},
  \bibinfo{author}{\bibfnamefont{T.}~\bibnamefont{Duan}},
  \bibinfo{author}{\bibfnamefont{X.}~\bibnamefont{Wang}},
  \bibinfo{author}{\bibfnamefont{J.}~\bibnamefont{Huang}}, \bibnamefont{and}
  \bibinfo{author}{\bibfnamefont{T.}~\bibnamefont{Kang}},
  \bibinfo{journal}{Nucleic Acids Res.} \textbf{\bibinfo{volume}{43}},
  \bibinfo{pages}{7878} (\bibinfo{year}{2015}).

\bibitem[{\citenamefont{Croft et~al.}(2018)\citenamefont{Croft, Bolderson,
  Adams, El-Kamand, Kariawasam, Cubeddu, Gamsjaeger, and
  Richard}}]{2018-SCDB-Croft-Richard}
\bibinfo{author}{\bibfnamefont{L.}~\bibnamefont{Croft}},
  \bibinfo{author}{\bibfnamefont{E.}~\bibnamefont{Bolderson}},
  \bibinfo{author}{\bibfnamefont{M.}~\bibnamefont{Adams}},
  \bibinfo{author}{\bibfnamefont{S.}~\bibnamefont{El-Kamand}},
  \bibinfo{author}{\bibfnamefont{R.}~\bibnamefont{Kariawasam}},
  \bibinfo{author}{\bibfnamefont{L.}~\bibnamefont{Cubeddu}},
  \bibinfo{author}{\bibfnamefont{R.}~\bibnamefont{Gamsjaeger}},
  \bibnamefont{and} \bibinfo{author}{\bibfnamefont{D.}~\bibnamefont{Richard}},
  \bibinfo{journal}{Semin. Cell. Dev. Biol.}  (\bibinfo{year}{2018}),
  \urlprefix\url{https://doi.org/10.1016/j.semcdb.2018.03.014}.

\end{thebibliography}


\begin{thebibliography}{99}
%
\bibitem{SM-2016-PR-Manghi-Destainville} M. Manghi and N. Destainville, Phys. Rep. (2016), \textbf{631}, 1-41.
\bibitem{SM-2013-PRE-Dasanna-Manghi} A. Dasanna, N. Destainville, J. Palmeri and M. Manghi, Phys. Rev. E (2013), \textbf{87}, 052703.
\bibitem{SM-2015-JCP-Sicard-Manghi} F. Sicard, N. Destainville and M. Manghi, J. Chem. Phys. (2015), \textbf{142}, 034903.
\bibitem{SM-Hugel-PRL2005} T. Hugel, M. Rief, M. Seitz, H. E. Gaub, and R. Netz, Phys. Rev. Lett. (2005), \textbf{94}, 048301.
\bibitem{SM-Grosberg-AIP1994} A. Y. Grosberg and A. R. Khokhlov, Statistical Physics of Macromolecules (AIP, Melville, NY, 1994).
\bibitem{SM-Dasanna-EPL2012} A. K. Dasanna, N. Destainville, J. Palmeri, and M. Manghi, EuroPhys. Lett. (2012), \textbf{98}, 38002.
\bibitem{SM-Tinland-Macro1997} B. Tinland, A. Pluen, J. Sturm, and G. Weill, Macromolecules (1997), \textbf{30}, 5763.
\bibitem{SM-2004-JCP-Mielke-Benham} S.P. Mielke, W.H. Fink, V.V. Krishnan, N. Gr\o{}nbech-Jensen and C.J. Benham, J. Chem. Phys. (2004), \textbf{121}, 8104-8112.
\bibitem{SM-2005-JCP-Mielke-Benham} S.P. Mielke, N. Gr\o{}nbech-Jensen, V. Krishnan, W. Fink and C. Benham, J. Chem. Phys. (2005), \textbf{123}, 124911.
\bibitem{SM-2012-NJP-Olsen-Bohr} K. Olsen and J. Bohr, New Journal of Physics (2012), \textbf{14}, 023063.
\bibitem{SM-2004-MC-Cloutier-Widom} T. Cloutier and J. Widom, Mol. Cell (2004), \textbf{14}, 355-362.
\bibitem{SM-2005-PNAS-Cloutier-Widom} T. Cloutier and J. Widom, Proc. Nat. Acad. Soc. U.S.A. (2005), \textbf{102}, 3645-3650.
\bibitem{SM-2010-NAR-Bond-Maher} L. Bond, J. Peters, N. Becker, J. Kahn and L. M. III, Nucleic Acids Research (2010), \textbf{38}, 8072-8082.
\bibitem{SM-2012-Science-Shibata-Dutta} Y. Shibata, P. Kumar, R. Layer, S. Willcox, J. Gagan, J. Griffith and A. Dutta, Science (2012), \textbf{336}, 82-86.
\bibitem{SM-2015-CR-Dillon-Dutta} L. Dillon, P. Kumar, Y. Shibata, Y.-H.Wang, S. Willcox, J. Griffith, Y. Pommier, S. Takeda and A. Dutta, Cell Rep. (2015), \textbf{11}, 1749-1759.
\bibitem{SM-2010-PRE-Sayar-etal} M. Sayar, B. Avsaroglu and A. Kabakcioglu, Phys. Rev. E (2010), \textbf{81}, 041916.
\bibitem{SM-2015-eLS-Bowater} R. Bowater, Supercoiled DNA: Structure, eLS. John Wiley \& Sons, Ltd: Chichester, 2015.
\bibitem{SM-2005-OxfordUniversityPress-Bates-Maxwell} A. Bates and A. Maxwell, DNA topology (2nd Ed.), Oxford University Press, UK, 2005.
\bibitem{SM-2008-JPCB-Trovato-Tozzini} F. Trovato and V. Tozzini, J. Phys. Chem. B (2008), \textbf{112}, 13197-13200.
\bibitem{SM-2015-MS-Higgins-Vologodskii} N. Higgins and A. Vologodskii, Microbiol. Spectr. (2015), \textbf{3}, 1-49.
\bibitem{SM-2015-NC-Irobalieva-Zechiedrich} R. Irobalieva, J. Fogg, D. C. Jr., T. Sutthibutpong, M. Chen, A. Barker, S. Ludtke, S. Harris, M. Schmid, W. Chiu and L. Zechiedrich, Nature Comm. (2015), \textbf{6}, 8440.
\bibitem{SM-2012-SM-Adamcik-Dietler} J. Adamcik, J.-H. Jeon, K. Karczewski, R. Metzler and G. Dietler, Soft Matter (2012), \textbf{8}, 8651.
\bibitem{SM-2015-PA-Marko} J. Marko, Physica A (2015), \textbf{418}, 126-153.
\bibitem{SM-Murtola-PCCP2009} T. Murtola, A. Bunker, I. Vattulainen, M. Deserno, and M. Karttunen, Phys. Chem. Chem. Phys. (2009), \textbf{11}, 1869.
\bibitem{SM-Bustamante-COSB2000} C. Bustamante, S. B. Smith, J. Liphardt, and D. Smith, Curr. Opin. Struct.
Biol. 10, 279 (2000).
\bibitem{SM-1986-JMB-White-Bauer} J. White and W. Bauer, J. Mol. Biol. (1986), \textbf{189}, 329.
\bibitem{SM-1995-PRE-Marko-Siggia} J. Marko and E. Siggia, Phys. Rev. E (1995), \textbf{52}, 2912-2938.
\bibitem{SM-2008-PRL-Barducci-Parrinello} A. Barducci, G. Bussi, and M. Parrinello, Phys. Rev. Lett. (2008), \textbf{100}, 020603.
\bibitem{SM-2014-PRL-Dama-Voth} J.F. Dama, M. Parrinello, and G.A. Voth, Phys. Rev. Lett. (2014), \textbf{112}, 240602.
\bibitem{SM-2014-CPC-Tribello-Bussi} G.A. Tribello, M. Bonomi, D. Branduardi, C. Camilloni and G. Bussi, Comput. Phys. Comm. (2014), \textbf{185}, 604-613.
\bibitem{SM-2009-JCC-Bonomi-Parrinello} M. Bonomi, A. Barducci and M. Parrinello, J Comput. Chem. (2009), \textbf{30}, 1615.
\bibitem{SM-2010-JCP-Xin-Hamelberg} Y. Xin and D. Hamelberg, J. Chem. Phys. (2010), \textbf{132}, 224101.
\bibitem{SM-2013-PRL-Tiwary-Parrinello} P. Tiwary and M. Parrinello, Phys. Rev. Lett. (2013), \textbf{111}, 230602.
\bibitem{SM-2014-JCTC-Salvalaglio-Parrinello} M. Salvalaglio, P. Tiwary and M. Parrinello, J. Chem. Theory Comput. (2014), \textbf{10}, 1420-1425.
\bibitem{SM-cranR} The Comprehensive R Archive Network. https://cran.r-project.org/ (accessed January 17, 2018).
\bibitem{SM-2017-JCP-Gimondi-Salvalaglio} I. Gimondi and M. Salvalaglio, J. Chem. Phys. \textbf{147}, 114502 (2017).
\bibitem{SM-2018-arXiv-Sicard} F. Sicard, arXiv:1803.03490 [cond-mat.stat-mech] (2018).
\bibitem{SM-2013-JCP-Chen-Xiao} C. Chen, Y. Huang, and Y. Xiao, J. Chem. Phys. \textbf{138}, 164122 (2013).
\bibitem{SM-2005-JPCM-Ambjornssson-Metzler} T. Ambjornssson and R. Metzler, J. Phys. Condens. Matter 17, S1841 (2005).
\end{thebibliography}

\pagebreak
\widetext
\begin{center}
\textbf{\large Control of DNA denaturation bubble nucleation to advance nano-biosensing
\vskip 0.3cm
			   Supporting Information}
\end{center}
\setcounter{equation}{0}
\setcounter{figure}{0}
\setcounter{table}{0}
\setcounter{page}{1}
\makeatletter
\renewcommand{\theequation}{S\arabic{equation}}
\renewcommand{\thefigure}{S\arabic{figure}}
\renewcommand{\thetable}{S\arabic{table}}
\renewcommand{\bibnumfmt}[1]{[S#1]}
\renewcommand{\citenumfont}[1]{S#1}
\newcommand{\sidebysidecaption}[4]{%
\RaggedRight%
  \begin{minipage}[t]{#1}
    \vspace*{0pt}
    #3
  \end{minipage}
  \hfill%
  \begin{minipage}[t]{#2}
    \vspace*{0pt}
    #4
\end{minipage}%
}
%
\section{Numerical model}
\begin{figure}[b]
\includegraphics[width=0.45 \textwidth, angle=-0]{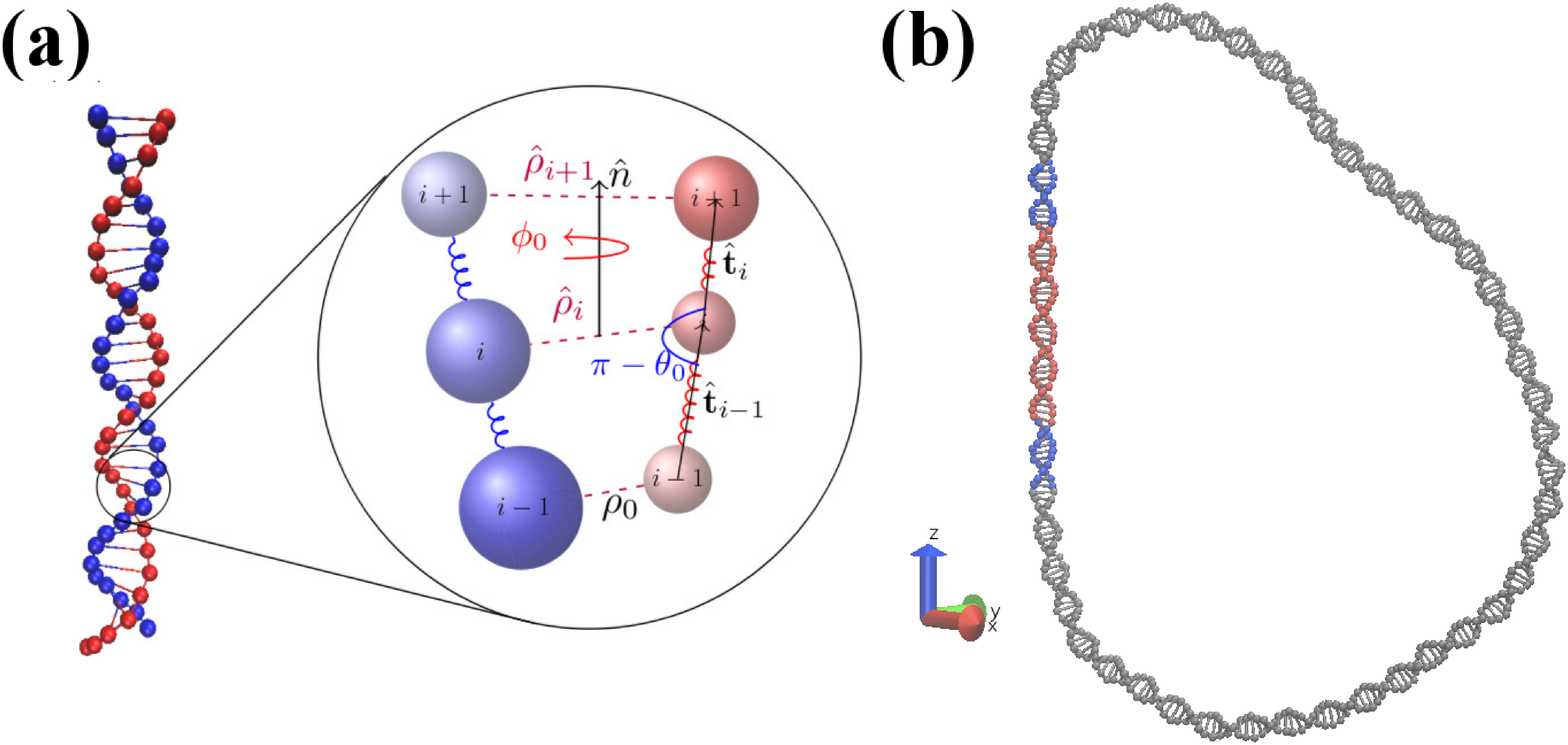}
 \caption{
 (a) Snapshot of an equilibrated double helix (from \cite{SM-2013-PRE-Dasanna-Manghi}). 
 The bending angle along each strand is  $\theta_0$, $\rho_0$ is the equilibrium base-pair 
 distance and $\hat{n}$ is the helical axis  around which twist is defined. The imposed equilibrium 
 twist between successive pairs is $\phi_0$.
 (b) Snapshot of the equilibrated circular dsDNA, $\textrm{cDNA}_0$, of size $N=300$ bps and pitch initially set 
 to $p=12.0$ bps. The AT-rich region of size $30$ bps (red color) 
 is delimited on each extremity by two sequences of 10 GC bps aligned arbitrarily along the Z-axis (blue color). 
 The molecule is closed by a circular GC region of size $(N-50)$ bps (grey color).  
}
\label{SM-fig1}
\end{figure}
To overcome the inherent limitations of atomistic simulations encountered at length- and time-scales  
of interest~\cite{SM-2016-PR-Manghi-Destainville}, we use the DNA model of Refs.~\onlinecite{SM-2013-PRE-Dasanna-Manghi,
SM-2015-JCP-Sicard-Manghi}, where the mesoscopic DNA model consists in two interacting bead-spring chains each made of N beads 
(of diameter $a=0.34$ nm) at position $\textbf{r}_i$, with a AT-rich region of $30$ bps clamped with a GC region of $N-30$ bps, 
as shown in Fig.~\ref{SM-fig1}. The Hamiltonian is $\mathcal{H}=\mathcal{H}_{el}^{(1)} + \mathcal{H}_{el}^{(2)} + \mathcal{H}_{tor} + \mathcal{H}_{int}$, where the first two contributions are elastic energies of the strands $j=1,2$, which include both stretching 
and bending energies
\begin{equation}
\mathcal{H}_{el}^{(j)} = \sum_{i=0}^{N-1} \frac{\kappa_s}{2}(r_{i,i+1}-a_\textrm{ref})^2 
+ \sum_{i=0}^{N-1}\frac{\kappa_\theta}{2}(\theta_i-\theta_\textrm{ref})^2.
\end{equation}
The stretching modulus, $a^2\beta_0 \kappa_s = 100$, is a compromise between numerical efficiency 
and experimental values~\cite{SM-Hugel-PRL2005}, where $\beta_0^{-1} = k_B T$ is the thermal energy,
$T = 300$ K is the room temperature, and $a_\textrm{ref}=0.357$ nm. The bending modulus is large, $\beta_0 \kappa_\theta = 600$, 
to maintain the angle between two consecutive tangent vectors along each strand $\theta_i$ to the 
fixed value $\theta_\textrm{ref} = 0.41$ rad. Each strand is thus modeled as a freely rotating chain (FRC)~\cite{SM-Grosberg-AIP1994}.
The third and fourth terms of $\mathcal{H}$ are the torsional energy and hydrogen-bonding interactions, respectively. 
The torsional energy is modeled by a harmonic potential
\begin{equation}
\mathcal{H}_{tor} = \sum_{i=0}^{N-1} \frac{\kappa_{\phi,i}}{2}(\phi_i-\phi_\textrm{ref})^2 ,
\end{equation}
where $\phi_i$ is defined as the angle between two consecutive base-pair vectors
$\brho_i \equiv \textbf{r}_i^{(1)}-\textbf{r}_i^{(2)}$ and  $\brho_{i+1}$ ($\phi_\textrm{ref} = 0.62$ rad). 
The stacking interaction between base pairs is modeled through a $\kappa_{\phi,i}$ that depends on 
the value of the \textit{bare} dsDNA torsional modulus $\kappa_\phi$, and the distances between complementary bases,
$\kappa_{\phi,i} = \kappa_\phi [1-f(\rho_i)f(\rho_{i+1})]$, where 
\begin{equation} 
f(\rho_i) = \frac{1}{2}\Big[1+\erf\Big(\frac{\rho_i -\rho_b}{\lambda'}\Big)\Big], 
\label{stacking}
\end{equation}
and $\rho_i =|\brho_i|$. Hence, $\kappa_{\phi,i} = \kappa_\phi$ in the dsDNA state and $\kappa_{\phi,i} = 0$ 
in the ssDNA one. The actual values in the dsDNA state after equilibration, $\kappa^*_{\phi,\rm ds}$, 
are however different from the prescribed values, $\kappa_{\phi}$, due to thermal fluctuations and non-linear potentials 
entering the Hamiltonian. 
The hydrogen-bonding interaction is modeled by a Morse potential
\begin{equation}
\mathcal{H}_{int} = \sum_{i=0}^{N-1} A (e^{-2\frac{\rho_i-\rho_\textrm{ref}}{\lambda}} -2e^{-\frac{\rho_i-\rho_\textrm{ref}}{\lambda}}) ,
\end{equation}
where $\rho_\textrm{ref}=1$ nm, $\lambda=0.2$ nm, and $\beta_0 A=8$ and $12$ for AT and GC bonding, respectively, 
as in Refs.~\onlinecite{SM-Dasanna-EPL2012, SM-2013-PRE-Dasanna-Manghi,SM-2015-JCP-Sicard-Manghi}.
The fitted values for the dsDNA persistence length and the pitch are $\ell_{\rm ds}\simeq160$~bps 
and $p = 12$~bps for the relevant range of $\beta_0\kappa_\phi$ we are interested in, which are comparable to 
the actual dsDNA values ($\ell_{\rm ds}\simeq150$~bps and $p= 10.4$~bps). The ssDNA persistence length 
is $\ell_{\rm ss} = 3.7$~nm, compatible with experimental measurement~\cite{SM-Tinland-Macro1997}, 
even though in the upper range of measured values.\\

The dsDNA minicircle is described by a circular helix where a helical line of radius $\alpha$ coils around 
a torus of radius $R$ in the $x-y$ plane~\cite{SM-2004-JCP-Mielke-Benham,SM-2005-JCP-Mielke-Benham,SM-2012-NJP-Olsen-Bohr}. 
The centers of the beads on each strand initially coincide with 
the surface of this torus in Cartesian space according to the equations
\begin{equation}
\left\{
\begin{aligned}
x_n^{(j)} &= \Big( \alpha \sin\Big(n\frac{2\pi}{p} + \psi^{(j)}\Big) + R \Big) \times \cos(n\theta) \\
y_n^{(j)} &= \Big( \alpha \sin\Big(n\frac{2\pi}{p} + \psi^{(j)}\Big) + R \Big) \times \sin(n\theta) \\
z_n^{(j)} &= \alpha \cos\Big(n\frac{2\pi}{p} + \psi^{(j)}\Big)
\end{aligned}
\right.
\end{equation}
with $x_n^{(j)}$, $y_n^{(j)}$ and $z_n^{(j)}$ the Cartesian coordinates of  bead $n$ on strand $j$. 
The parameter $\psi^{(1)}=0$ for the first strand and $\psi^{(1)}=\pi$ for the second strand. The cross-sectional 
radius $\alpha$ is set equal to half the equilibrium base-pair distance, $\rho_{\textrm{ref}} = 1$~nm, considered 
in previous work~\cite{SM-2013-PRE-Dasanna-Manghi,SM-2015-JCP-Sicard-Manghi}. 
The twist angle between two base-pairs is defined as $\phi =2\pi/p $, where $p$ is the DNA pitch, 
\textit{i.e.} the number of bps corresponding to one complete helix turn. For purposes of generating the initial 
conformations, the bending angle per axis segment between the centers of two consecutive bps 
is set initially at $\theta = 2\pi/N$.\\

In the following, we restrained our analysis to four different circular dsDNAs (cDNA) with different superhelical 
density, $\sigma$, but with a similar sequence of bps.
As shown in Table~\ref{SM-DNA-parameters}, the \textit{reference} pitchs, $p^{(th)}$, of $\textrm{cDNA}_0$, 
$\textrm{cDNA}_1$, $\textrm{cDNA}_2$ and $\textrm{cDNA}_3$ are initially set to $p^{(th)} = 12.0$, $12.1$, $12.3$, 
and $12.5$~bps, respectively.
The number of beads on each strand, $N$, is chosen so that the number of axis segment, 
$N/p$, be an integer, and $\ell_{\rm ds} < N < 400$ bps, as it is representative of the 
supercoiled DNA loops found in nature~\cite{SM-2004-MC-Cloutier-Widom,SM-2005-PNAS-Cloutier-Widom,SM-2010-NAR-Bond-Maher,
SM-2012-Science-Shibata-Dutta,SM-2015-CR-Dillon-Dutta}.
The superhelical densities, along with the sizes $N$ of the minicircles, were specifically chosen 
to tune the value of $\Delta Lk < 1$. Such specific design allowed us to control the interplay between twist and writhe 
during the formation of the \textit{long-lived} denaturation bubbles~\cite{SM-2010-PRE-Sayar-etal}.
Furthermore, to quantify the role of the boundary/closure conditions on the formation of the denaturation bubble, 
we considered a linear dsDNA of $N=50$ bps made of a similar AT-rich region of $30$ bps clamped by GC regions of $10$ bps 
on each extremity ($\ell\textrm{DNA}$ in Tab.~\ref{SM-DNA-parameters}).
\begin{table*}[t]
\begin{center}
\caption{Linear ($\ell$DNA) and circular (cDNA) dsDNA parameters for the set of initial configurations of sizes, $N$, 
and difference in linking number, $\Delta Lk$, considered throughout this study. The theoretical (\textit{th}) 
and numerical (\textit{num}) values obtained for the equilibrated dsDNA are given for the pitch, $p$, 
twist angle, $\phi$, writhe, $Wr$, and superhelical density, $\sigma$.}
\begin{tabular*}{0.9\textwidth}{@{\extracolsep{\fill}}ccccccccc}
  \hline\hline
  {} & $p^{(th)}$ (bps) & $N$ (bps) & $N/\ell_{\rm ds}$ & $\Delta Lk$ & $\phi_{\textrm{eq}}^{(th)}$ (rad) & $\phi_{\textrm{eq}}^{(num)}$ (rad) & $Wr_{\textrm{eq}}^{(num)}$ & $\sigma$ \\
  \hline\hline
  $\ell\textrm{DNA}$ & $12.0$ & $50$ & --- & $0$ & $0.524$ & $0.547 \pm 0.047$ & --- & --- \\
  $\textrm{cDNA}_0$ & $12.0$ &$300$ & $1.9$ & $0$ & $0.524$ & $0.528 \pm 0.050$ & $0.02 \pm 0.03$ & $0$ \\
  \hline
  $\textrm{cDNA}_{1a}$ & $12.1$ & $242$ & $1.5$ & $-0.17$ & $0.519$ & $0.527 \pm 0.051$ & $-0.09 \pm 0.06$ & $-0.008$ \\
  $\textrm{cDNA}_{1b}$ & $12.1$ & $363$ & $2.3$ & $-0.25$ & $0.519$ & $0.527 \pm 0.051$ & $-0.17 \pm 0.06$ & $-0.008$ \\
  \hline
  $\textrm{cDNA}_{2a}$ & $12.3$ & $246$ & $1.5$ & $-0.5$ & $0.511$ & $0.520 \pm 0.051$ & $-0.15 \pm 0.04$ & $-0.024$ \\
  $\textrm{cDNA}_{2b}$ & $12.3$ & $369$ & $2.3$ & $-0.75$ & $0.511$ & $0.523 \pm 0.052$ & $-0.38 \pm 0.05$ & $-0.024$ \\
  \hline
  $\textrm{cDNA}_{3a}$ & $12.5$ & $225$ & $1.4$ & $-0.75$ & $0.503$ & $0.519 \pm 0.052$ & $-0.37 \pm 0.06$ & $-0.04$ \\
  $\textrm{cDNA}_{3b}$ & $12.5$ & $250$ & $1.6$ & $-0.83$ & $0.503$ & $0.524 \pm 0.051$ & $-0.61 \pm 0.08$ & $-0.04$ \\
  \hline\hline
\end{tabular*}
\label{SM-DNA-parameters}
\end{center}
\end{table*}

To allow comparison of the degree of supercoiling in molecules of different sizes, we normalize measurements 
of supercoiling with the use of the superhelical density~\cite{SM-2010-PRE-Sayar-etal,SM-2015-eLS-Bowater}
\begin{equation}
\sigma = \frac{Lk - Lk^0}{Lk^0} = \frac{\Delta Lk}{Lk^0}\,,
\end{equation}
where $Lk$ represents the linking numbers of the cDNA molecule, \textit{i.e.} the number of times 
one backbone strand \textit{links through} the circle formed by the other~\cite{SM-2010-PRE-Sayar-etal,SM-2015-eLS-Bowater}, 
and $Lk^0$ is defined as $Lk^0 = N/p_0$ for any DNA molecule, with $p_0 = 12.0$ bps the equilibrium pitch measured 
in the linear state. 
For instance, natural circular DNA molecules, such as bacterial plasmids, vary widely in size, but, when isolated 
\textit{in vitro}, the majority have values for $ \sigma \leq -0.03$~\cite{SM-2005-OxfordUniversityPress-Bates-Maxwell,
SM-2008-JPCB-Trovato-Tozzini,SM-2015-eLS-Bowater,SM-2015-MS-Higgins-Vologodskii}.
$Lk$ is a topological property of circular DNA that does not depend on its particular 
conformation~\cite{SM-2012-NJP-Olsen-Bohr,SM-2012-SM-Adamcik-Dietler}, and  obeys the relation
\begin{equation}
\label{Lk-definition}
Lk = Tw + Wr \, ,
\end{equation}
where $Tw$ represents the helical twist (the number of times either backbone winds around the helix axis), 
and $Wr$ represents the writhe, or degree of supercoiling (the number of signed crossing of the helix axis 
in planar projection, averaged over all projection directions). Although $Lk$ is a topological invariant integer, 
$Wr$ and $Tw$ are not and depend on geometry~\cite{SM-2015-PA-Marko}. For a given molecule, the superhelical stress 
produced by deviations of $Lk$ from $Lk^0$ is accomodated by changes in $Tw$, $Wr$, or both, following
\begin{equation}
\label{Lk-deviation}
\Delta Lk = (Lk-Lk^0) = \Delta Tw + \Delta Wr \, .
\end{equation}
Here, $\Delta Tw$ corresponds to localized, sequence-dependent twist deformations such as strand separation or 
double-helical structure transitions. $\Delta Wr$ corresponds to bent (supercoiling) deformations~\cite{SM-2015-PA-Marko}.
\begin{table}[t]
\begin{center}
\caption{Linear ($\ell$DNA) and circular (cDNA) dsDNA characteristics obtained within the accelerated dynamics framework. 
The parameters $\rho_{\textrm{max}}^{bub}$, $\phi_{\textrm{min}}^{bub}$, 
$N^{bub}_{\textrm{av}}$ and $N^{bub}_{\textrm{max}}$ correspond to the location of the nucleation basin 
in the free energy surfaces reconstructed in Fig.~1d in the main text, and the average and maximal number of opened base-pairs 
in the denaturation bubble, respectively. The uncertainties on $\rho_{\textrm{max}}^{bub}$ and $\phi_{\textrm{min}}^{bub}$ 
are measured from the isosurface delimited within $1~k_B T$ from the free energy minimum in the free energy surfaces 
reconstructed in Fig.~1d in the main text.
}
\begin{tabular*}{0.8\textwidth}{@{\extracolsep{\fill}}ccccccc}
  \hline\hline
  {} & $p^{(th)}$ (bps) & $\Delta Lk$ & $N/\ell_{\rm ds}$ & $\rho_{\textrm{max}}^{bub}$ (nm) & $\phi_{\textrm{min}}^{bub}$ (rad) & $N^{bub}_{\textrm{av}}$ ($N^{bub}_{\textrm{max}}$)  \\
  \hline \hline
  $\ell\textrm{DNA}$ & $12.0$ & $0$ & --- & $1.8 \pm 0.2$ & $0.15 \pm 0.05$ & $9 \pm 3$ ($16$) \\
  $\textrm{cDNA}_0$ & $12.0$ & $0$ & $1.9$ & $1.7 \pm 0.1$ & $0.17 \pm 0.05$ & $8 \pm 2$ ($14$)  \\
   \hline
   $\textrm{cDNA}_{1a}$ & $12.1$ & $-0.17$ & $1.5$ & $1.8 \pm 0.1$ & $0.15 \pm 0.04$ & $8 \pm 2$ ($16$)  \\
  $\textrm{cDNA}_{1b}$ & $12.1$ & $-0.25$ & $2.3$ & $1.8 \pm 0.2$ & $0.15 \pm 0.05$ & $9 \pm 2$ ($18$) \\
   \hline
  $\textrm{cDNA}_{2a}$ & $12.3$ & $-0.5$ & $1.5$ & $2.4 \pm 0.2$ & $0.06 \pm 0.05$ & $12 \pm 2$ ($20$)  \\
  $\textrm{cDNA}_{2b}$ & $12.3$ & $-0.75$ & $2.3$ & $2.3 \pm 0.5$ & $0.08 \pm 0.04$ & $12 \pm 3$ ($22$) \\
   \hline
  $\textrm{cDNA}_{3a}$ & $12.5$ & $-0.75$ & $1.4$ & $3.0 \pm 0.5$ & $-0.01 \pm 0.06$ & $14 \pm 3$ ($22$)  \\
  $\textrm{cDNA}_{3b}$ & $12.5$ & $-0.83$ &  $1.6$ & $3.1 \pm 0.5$ & $-0.02 \pm 0.06$ & $15 \pm 3$ ($26$)  \\
  \hline\hline
\end{tabular*}
\label{SM-DNA-FES}
\end{center}
\end{table}
\subsection{MD simulation}
The evolution of the system is governed by Brownian dynamics, \textit{i.e.} simulations based upon numerical 
integration of the overdamped Langevin equation~\cite{SM-2004-JCP-Mielke-Benham,SM-2005-JCP-Mielke-Benham,
SM-2013-PRE-Dasanna-Manghi,SM-2015-JCP-Sicard-Manghi}. 
The evolution of $\textbf{r}_i(t)$ is governed by the overdamped Langevin equation, 
integrated using a Euler's scheme,
\begin{equation}
\zeta \frac{d\textbf{r}_i}{dt} = -\nabla_{\textbf{r}_i}\mathcal{H}({\textbf{r}_j}) + \mathbf{\xi}(t) ,
\end{equation}
where $\zeta=3\pi\eta a$ is the friction coefficient for each bead of diameter $a$ with 
$\eta=10^{-3}$ Pa.s the water viscosity. 
The diffusion coefficient, $D_\textrm{diff} \equiv k_BT/3\pi\eta a$, thus takes into account 
the level of coarse-graining of the mesoscopic model involved in the kinetics associated 
to the smoothed free energy landscape~\cite{SM-Murtola-PCCP2009}. 
The random force of zero mean $\mathbf{\xi}_i(t)$ obeys 
the fluctuation-dissipation relation $\langle \mathbf{\xi}_i(t).\mathbf{\xi}_i(t')\rangle =6k_BT\zeta\delta_{ij}\delta(t-t')$. 
Lengths and energies are made dimensionless in the units of $a=0.34$ nm and $k_BT$, respectively. 
The dimensionless time step is $\delta\tau = \delta t k_B T/(a^2\zeta)$, set to $5 \times 10^{-4}$ 
($\delta t=0.045$ ps) for sufficient accuracy~\cite{SM-Dasanna-EPL2012,SM-2013-PRE-Dasanna-Manghi,SM-2015-JCP-Sicard-Manghi}. 
This set of parameters induces zipping velocities $v \approx 0.2-2$ bp/ns, compatible with experimental 
measurements~\cite{SM-Bustamante-COSB2000}.\\

The \textit{initial} DNA state was first constrained in a plane to relax its geometrical parameters, such as stretching, 
bending along a single strand, and twisting, keeping the writhe of the system null. The geometrical constraint was then released, 
so that the system relaxed its linking number between helical twist $Tw$ and writhe $Wr$, 
as described in Eq.~\ref{Lk-definition} and reported in Tab.~\ref{SM-DNA-parameters}. 
Following the work of Mielke et al.~\cite{SM-2004-JCP-Mielke-Benham}, the latter dynamical quantity can be derived from 
the discretization of the White's integral expression~\cite{SM-1986-JMB-White-Bauer,SM-1995-PRE-Marko-Siggia},
\begin{equation}
\label{WhiteIntegral}
4\pi Wr = \sum_j \sum_{i\neq j} \Big( (\textbf{r}_{j+1}-\textbf{r}_{j}) \times (\textbf{r}_{i+1}-\textbf{r}_{i}) \Big)
. \frac{(\textbf{r}_{j}-\textbf{r}_{i})}{|\textbf{r}_{j}-\textbf{r}_{i}|^3} \,.
\end{equation}
The dot product in Eq.~\ref{WhiteIntegral} determines the magnitude of relative nonplanar bending of the segments 
of the helix axis defined by the pair of axis vectors, $(\textbf{r}_{i+1}-\textbf{r}_{i})$ 
and $(\textbf{r}_{j+1}-\textbf{r}_{j})$, with $\textbf{r}_{i} \equiv (\textbf{r}_i^{(1)}+\textbf{r}_i^{(2)})/2$. 
The instantaneous writhe of each substructure is found by summing over all pairs.
\begin{figure}[b]
\includegraphics[width=0.5 \textwidth, angle=-0]{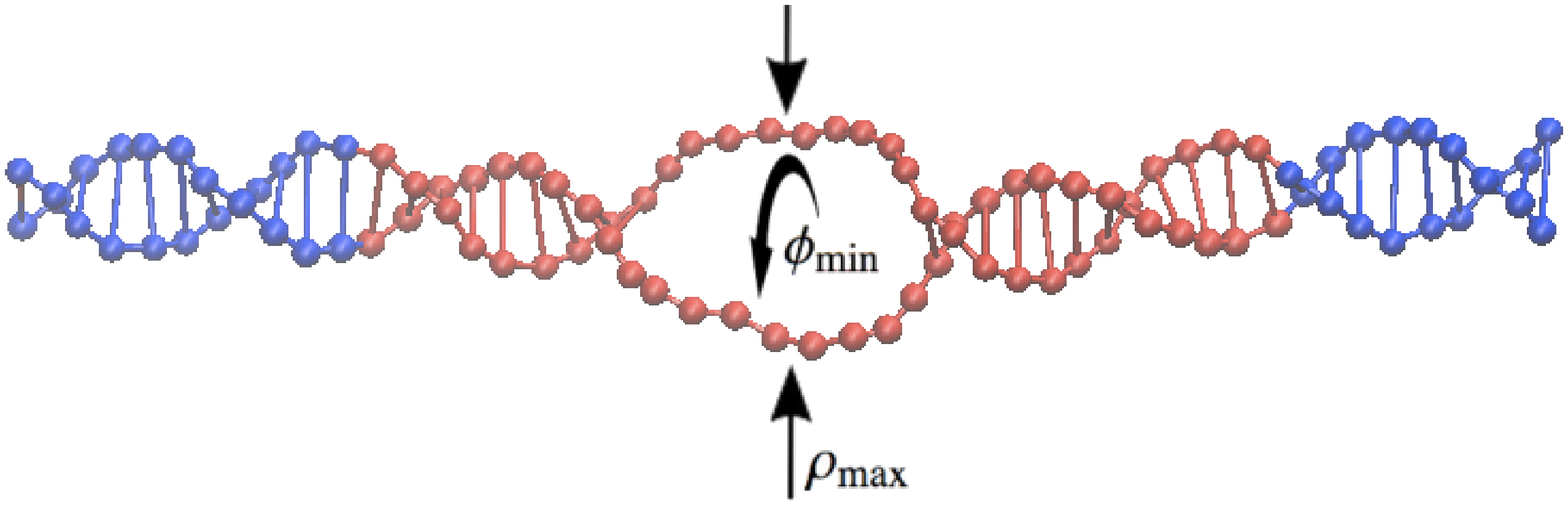}
 \caption{
Snapshot of the linear dsDNA, $\ell DNA$, defined in Table~\ref{SM-DNA-parameters}, when the 
\textit{long-lived} denaturation bubble is formed. The AT-rich region of size $30$ bps (red color) 
is clamped by two sequences of 10 GC bps on each extremity, which are aligned arbitrarily along the Z-axis (blue color). 
The two collective variable considered in the metaD simulation are shown: the maximal distance between paired bases, 
$\rho_{\textrm{max}}$, and the minimal twist angle between successive bps, $\phi_{\textrm{min}}$, as in Fig. 1b in the main text.
}
\label{fig2}
\end{figure}

\section{Biased MD simulation}
\textbf{Thermodynamic properties.} The well-tempered variant of the metadynamics (WT-metaD) 
enhanced sampling technique~\cite{SM-2008-PRL-Barducci-Parrinello,SM-2014-PRL-Dama-Voth} was implemented with 
the coarse-grained (CG) Brownian simulations 
of the circular and linear dsDNA, and performed using the version 2.3 of the plugin 
for free energy calculation, named PLUMED~\cite{SM-2014-CPC-Tribello-Bussi}. WT-metaD enhances the sampling of the 
conformational space of the system along a few selected degrees of freedom, 
named \textit{collective variables} (CVs), and reconstructs the equilibrium probability distribution, 
and thus the free energy landscape, as a function of these CVs. 
As shown in Fig.~\ref{fig2}, and already discussed in previous work~\cite{SM-2015-JCP-Sicard-Manghi}, we considered 
the width $\rho_{\textrm{max}}$ of the denaturation bubble, \textit{i.e.} the maximal distance between paired bases, 
as CV to bias the dynamics. We also choose to follow the evolution of the minimal twist angle 
inside the bubble, $\phi_{\textrm{min}} = \textrm{min}_{i \in \textrm{bubble}}\phi_i$, 
where $\phi_i$ is defined as the angle between two consecutive base-pair vectors $\brho_i$ and $\brho_{i+1}$.

According to the algorithm introduced by Barducci et al.~\cite{SM-2008-PRL-Barducci-Parrinello}, 
a Gaussian bias 
potential is deposited every $\tau_G$ with height $\omega = \omega_0 e^{-V(s,t)/(f-1)T}$, where $s$ 
is the CV, $\omega_0$ is the initial height, $T$ is the temperature of the simulation, $V(s,t)$ 
the metadynamics time-dependent bias,
\begin{equation}
\label{metaD-potential}
V(s,t) = \omega \sum_{t'<t}\exp\Big[ -\frac{(s(t)-s(t'))^2}{2\delta^2}\Big]\, ,
\end{equation}
and $f\equiv(T+\Delta T)/T$ is the bias factor with $\Delta T$ a parameter with the dimension of a temperature.
The resolution of the recovered free energy landscape is determined by the width of the Gaussian $\delta$. 
We put a restraint \textit{wall} potential at specific values of $\rho_{\textrm{max}}$ to prevent the system to escape 
from the metastable state. We checked that a slight change in the position of the \textit{wall} did not change significantly 
the results, particularly the positions of the local minimum and the saddle point, as well as the barrier height. 
The simulations are run until the free energy profile does not change more than $2~k_B T$ in the last $100$ ns. 
To further control the error of the reconstructed landscape, we performed $3$ runs of WT-metaD for each DNA minicircle.
The other observables are reconstructed afterwards using the \textit{reweighting technique} 
of Bonomi et al.~\cite{SM-2009-JCC-Bonomi-Parrinello}. 
The different sets of values considered in the WT-metaD simulations are given in Table~\ref{DNA-metaD}.\\

\textbf{Dynamical properties.} Building on the accelerated dynamics framework of Hamelberg et al.~\cite{SM-2010-JCP-Xin-Hamelberg}
and more recently Tiwary et al.~\cite{SM-2013-PRL-Tiwary-Parrinello,SM-2014-JCTC-Salvalaglio-Parrinello}, we extended the Metadynamics 
scope to estimate the mean transition times between the metastable (bubble) 
and the equilibrium (closed) states observed in the circular and linear dsDNA. WT-metaD was performed using 
the width $\rho_{\textrm{max}}$ of the denaturation bubble as CV.
We denote by $\tau$, the mean transition time over the barrier from the metastable state to the closed state, 
and by $\tau_M$, the mean transition time for the metadynamics run. The latter changes as the simulation 
progresses and is linked to the former through the acceleration factor 
$\alpha(t) \equiv \langle e^{\beta V(s,t)} \rangle_M = \tau/\tau_M(t)$, 
where the angular brackets $\langle \dots \rangle_M$ denote an average over a metadynamics run confined to the metastable basin, 
and $V(s,t)$ is the metadynamics time-dependent bias. To satisfy the main validity criterions, \textit{ie.} 
1) to consider a set of CVs able to distinguish between the different metastable states~\cite{SM-2014-JCTC-Salvalaglio-Parrinello}, 
and 2) to avoid depositing bias in the Transition State region~\cite{SM-2013-PRL-Tiwary-Parrinello}, we check that 
the statistics of transition times follows a Poisson distribution, and increase the time lag between two successive Gaussian depositions $\tau_G = \tau_G^{(\textrm{dyn})}$, as indicated in Tab.~\ref{DNA-metaD}. 
We performed several WT-metaD simulations and stop the simulations when the crossing 
of the barrier and the Gaussian deposition occur unlikely at the same time. 
To assess the reliability of the biased simulations, we checked that no bias potential 
was added to the transition state region during the WT-metaD simulations~\cite{SM-2014-JCTC-Salvalaglio-Parrinello}.          
We also performed statistical analysis of the distribution of transition times. 
We performed a two-sample Kolmogorov-Smirnov (KS) test, which does not require a priori knowledge 
of the underlying distribution~\cite{SM-2014-JCTC-Salvalaglio-Parrinello}. 
We tested the null hypothesis that the sample of transition times extracted from the metaD simulations and a large 
sample of times randomly generated according to the theoretical exponential distribution reflect the same 
underlying distribution. The null hypothesis is conventionally rejected if the $p$-value $< 0.05$. The KS test has been 
performed as implemented in the software cran-R~\cite{SM-cranR}.
\begin{table*}[t]
\begin{center}
\caption{Linear ($\ell$DNA) and circular (cDNA) dsDNA parameters considered in the accelerated dynamics framework. 
$\delta$ and $\omega_0$ refer to the width and the initial height of the Gaussian 
potentials, respectively. $f^{(\textrm{therm})}$, $\textrm{wall}^{(\textrm{therm})}$, 
$\tau_G^{(\textrm{therm})}$, and $f^{(\textrm{dyn})}$, $\textrm{wall}^{(\textrm{dyn})}$, $\tau_G^{(\textrm{dyn})}$, 
correspond to the bias factor, the location of the restraint \textit{wall} potential applied on $\rho_{\textrm{max}}$ 
and the bias deposition time in the metadynamics simulations dedicated to the reconstruction of 
the free energy landscape and the determination of the transition rates, respectively. 
The symbol (---) means no wall was considered.}
\begin{tabular*}{0.9\textwidth}{@{\extracolsep{\fill}}ccccccccc}
  \hline\hline
  {} & $\delta$ (nm) & $\omega_0$ (kJ/mol) & $\tau_G^{(\textrm{therm})}$ (ps) &  $\tau_G^{(\textrm{dyn})}$ (ps) & $f^{(\textrm{therm})}$ & $f^{(\textrm{dyn})}$ & $\textrm{wall}^{(\textrm{therm})}$ (nm) & $\textrm{wall}^{(\textrm{dyn})}$ (nm) \\
  \hline\hline
  $\ell\textrm{DNA}$ & $0.034$ & $3$ & $25$ & $700$ & $6$ & $3$ & $4.0$ & --- \\
  $\textrm{cDNA}_0$ & $0.034$ & $3$ & $25$ & $700$ & $6$ & $3$ & --- & --- \\
  $\textrm{cDNA}_{1a/1b}$ & $0.034$ & $3$ & $25$ & $700$ & $6$ & $3$ & --- & --- \\
  $\textrm{cDNA}_{2a/2b}$ & $0.034$ & $3$ & $25$ & $700$ & $20$ & $3$ & $4.0$ & --- \\
  $\textrm{cDNA}_{3a/3b}$ & $0.034$ & $3$ & $25$ & $700$ & $25$ & $3$ & $4.0$ & --- \\
  \hline\hline
\end{tabular*}
\label{DNA-metaD}
\end{center}
\end{table*}
\\

\begin{figure}[t]
\includegraphics[width=0.8 \textwidth, angle=-0]{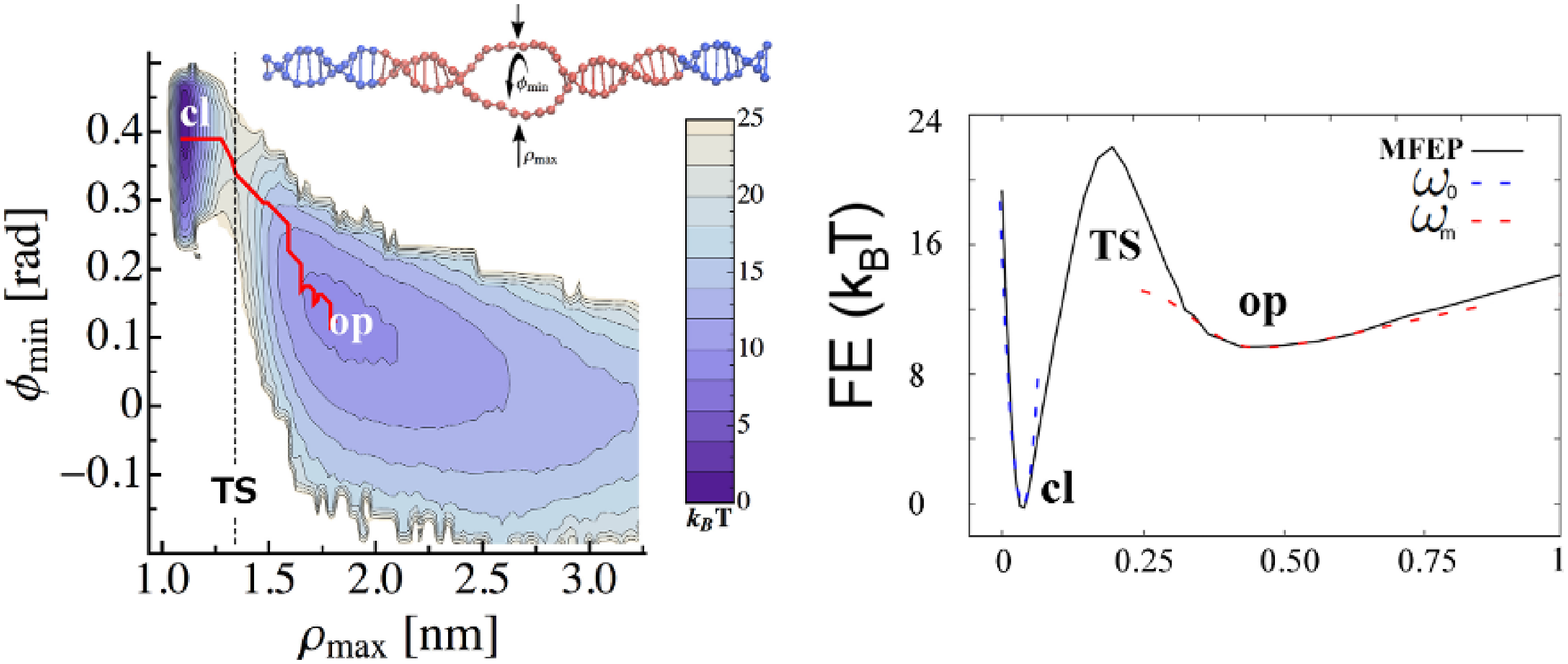}
 \caption{
\textbf{Left panel} Free energy surface associated with the linear DNA bubble closure/nucleation mechanism projected 
 along the maximal distance between paired bases $\rho_{\textrm{max}}$ and the minimal twist angle between successive bps, 
 $\phi_{\textrm{min}}$ (see inset).  The contour lines are every two $k_B T$. The two stables basins associated with the opened (op) and closed (cl) states of the DNA bubble and the typical minimal free energy path (MFEP) obtained within the steepest 
 descent framework~\cite{SM-2013-JCP-Chen-Xiao} are shown (red). 
\textbf{Right panel} Free energy of the DNA bubble as a function of the progression along the typical MFEP (normalized to unity) 
obtained within the steepest descent framework~\cite{SM-2013-JCP-Chen-Xiao}. The nonlinear least-squares Marquardt-Levenberg 
algorithm was implemented to fit the parameters $\omega_{op}$ and $\omega_{cl}$, measured in the opened and closed DNA states, 
respectively. In addition to the steepest descent framework used for $\rho_{\textrm{max}} \leq \rho_{\textrm{max}}^{bub}$, 
we considered the slowest evolution of the slope of the free energy path for $\rho_{\textrm{max}} > \rho_{\textrm{max}}^{bub}$ 
to reconstruct fully the MFEP.
}
\label{SM-FE-fitting}
\end{figure}

Considering the recent development of Sicard~\cite{SM-2018-arXiv-Sicard} approaching the issue of complex system where configurational 
entropy is competing with energy, we extended the metadynamics scope discussed above to assess the characteristic 
times associated with the opening and closure of the denaturation bubbles when their direct numerical estimation 
was not feasible. To do so, we computed the ratio between the rates associated with the transition between 
the two free energy basins associated with the closed and opened dsDNA states, $\mathcal{B}_{cl}$ and $\mathcal{B}_{op}$, 
respectively, reconstructed in the free energy surfaces shown in Fig.1d in the main text: 
\begin{equation}
\label{SM-KTfinal}
\frac{k_{cl}}{k_{op}} = e^{\Delta S^{\textrm{conf}}_{0}/k_B} ~ \frac{\omega_{cl}}{\omega_{op}} ~ \frac{\gamma_{op}}{\gamma_{cl}} 
 ~ e^{-\Delta F_0/k_B T} \,.
\end{equation}
In Eq.~\ref{SM-KTfinal}, $\omega_{op}$ and $\omega_{op}$ represent the effective stiffness of the free energy well associated 
with the opened and closed dsDNA states, respectively (as depicted in Fig.~\ref{SM-FE-fitting} for the linear DNA). To account for the asymmetric nature of the free energy landscape in the free energy basins, skew-Gaussian fitting of the minimal free energy path (MFEP) 
was considered as described in the work of Sicard~\cite{SM-2018-arXiv-Sicard}. The respectives values are reported in Tab.~\ref{SM-DNA-rates}.
The difference in configurational entropy, $\Delta S^{\textrm{conf}}_{0}$ was assessed as
\begin{equation}
\label{SM-Sconf-metaD}
 -T\Delta S^{\textrm{conf}}_{0} = \Delta F_0 - \Delta F^*_0 \,,
\end{equation}
with $\Delta F_0$ the free energy of formation between the two free energy basins associated with the closed and opened dsDNA states
measured along the MFEP depicted in the main text, and $\Delta F^*_0$ the algebric values of the free energy of formation 
taking into account the entropic contribution to the free energy basins. The later term was defined in terms of the probability 
distribution of the CVs~\cite{SM-2017-JCP-Gimondi-Salvalaglio,SM-2018-arXiv-Sicard}:
\begin{equation}
\label{SM-FE-entropy}
\Delta F^{*}_{0} = -k_B T ~\log \Big( \frac{P_{cl}}{P_{op}}\Big) \,.
\end{equation}
In Eq.~\ref{SM-FE-entropy}, $P_{cl}$ and $P_{op}$ are the probabilities of the closed and opened DNA states, respectively. 
The probability of each state is computed as the integral of the distribution within the energy basin, 
$\mathcal{B}$, it occupies on the CV-space,
\begin{equation}
 P_i = \iint_{(\rho_{\textrm{max}},\phi_{\textrm{min}}) \in \mathcal{B}_\textrm{i}} 
 f(\rho_{\textrm{max}},\phi_{\textrm{min}}) ~d\rho_{\textrm{max}} ~d\phi_{\textrm{min}} \,,
 \label{SM-Proba-FE}
\end{equation}
where $f$ is the joint probability density distribution function associated with the system FE.
We considered the successive isosurfaces depicted in Fig.~1b in the main text as integration domains.
Finally, considering the Rouse model valid for flexible polymer chain~\cite{SM-2005-JPCM-Ambjornssson-Metzler}, the effective friction coefficients 
in Eq.~\ref{SM-KTfinal} depend on the number of opened bps, $N_{bub}$, in the DNA bubble reported in Tab~\ref{SM-DNA-FES}. 
The typical size observed in the simulations yields the relation $\gamma_{op}/\gamma_{cl} \approx N_{bub}$ between 
the effective frictions in Eq.~\ref{SM-KTfinal}. The results are reported in Tab.~\ref{SM-DNA-rates}.

\section{Quantitative assessment of misalignment of the dsDNA arms}
In our extensive simlations, we restrained our analysis to the case where the sequence of $10$ GC bps delimiting 
the AT-rich region were forced to be aligned. This was intended to dissociate the bending and 
twist contributions in the nucleation and closure mechanism. However, more realistic approach 
would necessarily take into account some relative misalignment of the sequences on both sides 
of the AT-rich region, at least during the initiation stage of the denaturation bubble nucleation/closure.
The bending contribution can be assessed analytically by modeling the denaturation bubble 
as a single rotating joint because the typical bubble length $N_{bub}\sim 10$~bps is on the order of the ssDNA persistence 
length $\ell_p^{\rm ss} \simeq 4$~nm. Denoting $\kappa$ the joint bending rigidity, the denaturation bubble's state 
can be characterized by the angle $\theta$ and energy $\kappa (1-\cos \theta)$, and the associated partition function is 
\begin{eqnarray}
Z_{\kappa} & = & \int_0^{\pi} {\rm d}\theta \, 2\pi \sin \theta \, e^{-\beta \kappa(1-\cos \theta)} \nonumber \\
 & = & 4 \pi \, e^{-\beta \kappa} \, \frac{\sinh \beta \kappa}{\beta \kappa}.
\end{eqnarray}
As compared to the case where the arms are forced to be aligned, the free energy gain in the unconstrained case is 
$\Delta F_{\kappa} = -k_{\rm B}T \ln Z_{\kappa}$. 
The value of $\kappa$ in the present case  is difficult to evaluate because the joint is composed of several base-pairs. 
It can be estimated in a rough approximation as $\kappa \approx 2 \kappa_{\rm ss}/N_{bub}$ because there are two strands 
in the bubble of length $N_{bub}$~bps. If $\ell_p^{\rm ss} \simeq 4$~nm and $N_{bub} \sim 10$ bps, one gets $\beta \kappa \approx 2$ 
and $\Delta F_{\kappa} \simeq -1.2 \, k_{\rm B}T$. 
In all cases, the free energy increase due to arm alignment is lower than 
$\lim_{\kappa \rightarrow 0} |\Delta F_{\kappa} |= k_{\rm B}T \ln 4 \pi \simeq 2.5 \, k_{\rm B}T$, in agreement 
with the result reported in the main text for $\ell$DNA.

\begin{table}
\begin{center}
\caption{Linear ($\ell$DNA) and circular (cDNA) dsDNA parameters derived from the accelerated dynamics simulations.
The parameters $\omega_{op}$, $\omega_{cl}$ represent the effective stiffness of the free energy well associated 
with the opened and closed dsDNA states, respectively, as depicted in Fig.~\ref{SM-FE-fitting}. $k_{op}/k_{cl}$ 
is the ratio of the transition rates associated with the opening (op) and closure (cl) of the \textit{long-lived} 
DNA denaturation bubble. The characteristic times derived from the accelerated dynamics simulations are given along 
with the respective $p$-value obtained from the Kolmogorov-Smirnov test. The symbol (---) indicates that Eq.~\ref{SM-KTfinal} 
was considered to determine the characteristic time.}
\begin{tabular*}{0.95\textwidth}{@{\extracolsep{\fill}}cccccccc}
  \hline\hline
  {} & $\omega_{op}$ & $\omega_{cl}$  & $k_{op}/k_{cl}$ & $\tau_{op}$ & $p$-value & $\tau_{cl}$ & $p$-value \\
  \hline \hline
  $\ell\textrm{DNA}$ & $64.2 \pm 2.1$ & $5.3 \pm 0.2$ &  $(1.5 \pm 0.6)\times 10^{-3}$ & $(67 \pm 8)$ ms & $0.65$ & $(121 \pm 12)~\mu$s &  $0.86$ \\
  $\textrm{cDNA}_0$ & $62.7 \pm 7.6$ & $9.3 \pm 0.7$ &  $(4.5 \pm 3.6)\times 10^{-4}$ & $(51 \pm 3)$ ms & $0.80$ & $(17 \pm 2)~\mu$s & $0.52$ \\
   \hline
   $\textrm{cDNA}_{1a}$ & $60.3 \pm 1.6$ & $4.5 \pm 0.7$ & $(8.2 \pm 6.4)\times 10^{-2}$ & $(10.4 \pm 0.6)$ ms & $0.98$ & $(1.7 \pm 0.3)$ ms & $0.62$\\
   $\textrm{cDNA}_{1b}$ & $56.9 \pm 1.3$ & $5.0 \pm 0.3$ & $(7.0 \pm 4.8)\times 10^{-3}$ & $(16.5 \pm 0.7)$ ms & $0.82$ & $(0.33 \pm 0.02)$ ms & $0.79$ \\
   \hline
  $\textrm{cDNA}_{2a}$ & $3.7 \pm 0.2$ & $69.5 \pm 3.1$ &  $(1.1 \pm 0.3)\times 10^6$ & $(4.9 \pm 0.6)$ ms & $0.71$ & $(90 \pm 30)$ min & --- \\
  $\textrm{cDNA}_{2b}$ & $3.2 \pm 0.2$ & $78.2 \pm 5.5$ & $(2.6 \pm 1.9)\times 10^4$ & $(5.9 \pm 0.5)$ ms & $0.66$ & $(110 \pm 90)$ s & --- \\
   \hline
  $\textrm{cDNA}_{3a}$ & $4.2 \pm 0.3$ & $85.7 \pm 3.3$ & $(3.4 \pm 1.3)\times 10^6$ & $(7.2 \pm 0.6)$ ms & $0.71$ & $(6.8 \pm 3.2)$ h & --- \\
  $\textrm{cDNA}_{3b}$ & $2.3 \pm 0.1$ & $72.7 \pm 2.8$ & $(5.7 \pm 1.9)\times 10^6$ & $(14.2 \pm 1.0)$ ms & $0.93$ & $(22.5 \pm 9.0)$ h & --- \\
  \hline
\end{tabular*}
\label{SM-DNA-rates}
\end{center}
\end{table}

\end{document}